\documentclass[fleqn,usenatbib]{mnras}
\usepackage{ae,aecompl}
\usepackage[T1]{fontenc}
\usepackage[latin9]{inputenc}
\usepackage{color}
\usepackage{array}
\usepackage{refstyle}
\usepackage{booktabs}
\usepackage{mathtools}
\usepackage{multirow}
\usepackage{amsmath}
\usepackage{amssymb}
\usepackage{graphicx}
\usepackage[authoryear]{natbib}
\ifx\hypersetup\undefined
  \AtBeginDocument{%
    \hypersetup{unicode=true}
  }
\else
  \hypersetup{unicode=true}
\fi

\makeatletter

\AtBeginDocument{\providecommand\secref[1]{\ref{sec:#1}}}
\AtBeginDocument{\providecommand\eqref[1]{\ref{eq:#1}}}
\AtBeginDocument{\providecommand\Tabref[1]{\ref{Tab:#1}}}
\AtBeginDocument{\providecommand\Eqref[1]{\ref{Eq:#1}}}
\AtBeginDocument{\providecommand\Figref[1]{\ref{Fig:#1}}}
\AtBeginDocument{\providecommand\figref[1]{\ref{fig:#1}}}
\AtBeginDocument{\providecommand\subsecref[1]{\ref{subsec:#1}}}
\AtBeginDocument{\providecommand\tabref[1]{\ref{tab:#1}}}
\providecommand{\tabularnewline}{\\}
\RS@ifundefined{subsecref}
  {\newref{subsec}{name = \RSsectxt}}
  {}
\RS@ifundefined{thmref}
  {\def\RSthmtxt{theorem~}\newref{thm}{name = \RSthmtxt}}
  {}
\RS@ifundefined{lemref}
  {\def\RSlemtxt{lemma~}\newref{lem}{name = \RSlemtxt}}
  {}

\usepackage{aecompl}
\usepackage{cuted}
\usepackage{caption}

\DeclareMathOperator\atanh{atanh}
\DeclareMathOperator\Li{Li}

\title[Katu]{Katu: a fast open-source full lepto-hadronic kinetic code suitable for Bayesian inference modeling of blazars}

\author[B. Jim\'enez-Fern\'andez et al.] {\
    Bruno Jim\'enez-Fern\'andez,$^{1}$\thanks{E-mail: B.Jimenez.Fernandez@bath.ac.uk}
H. J. van Eerten.$^{1}$\thanks{E-mail: hjve20@bath.ac.uk}
\\
$^{1}$Department of Physics, University of Bath, Claverton Down, Bath BA2 7AY, United Kingdom
}

\date{Accepted XXX\@. Received YYY; in original form ZZZ}

\pubyear{2021}

\makeatother

\begin{document}
\label{firstpage} \pagerange{\pageref{firstpage}--\pageref{lastpage}}
\maketitle
\begin{abstract}
In the modelling of blazars and other radiating plasma flows, correctly
evolving in time a series of kinetic equation coupled for different
particle species is a computationally challenging problem. In this
paper we introduce the code \texttt{Katu}, capable of simultaneously
evolving a large number of particle species including photons, leptons
and hadrons and their interactions. For each particle, the numerical
expressions for the kinetic equations are optimized for efficiency
and the code is written to easily allow for extensions and modifications.
Additionally, we tie \texttt{Katu} to two pieces of statistical software,
\texttt{emcee} and \texttt{pymultinest}, which makes a suite ready
to apply to blazars and extract relevant statistical information from
their electromagnetic (and neutrino, if applicable) flux. As an example,
we apply \texttt{Katu} to Mrk 421 and compare a leptonic and a leptohadronic
model. In this first ever direct Bayesian comparison between full
kinetic simulations of these models, we find substantial evidence
(Bayes factor $\sim7$) for the leptohadronic model purely from the
electromagnetic spectrum. The code is available online at \href{https://github.com/hveerten/katu}{https://github.com/hveerten/katu}
under the LGPL license\footnotemark.
\end{abstract}

\footnotetext{Alternatively, a mirror is maintained at \href{https://gitlab.com/Noxbru/Katu}{https://gitlab.com/Noxbru/Katu}}

\section{Introduction}

Simulating the behaviour of energetic particles both individually
and collectively is relevant in many fields of particle physics and
astrophysics. In the first case, understanding how a particle emits
and interacts with others has allowed us to build the theory of the
standard model of particle physics. In the second case, by putting
together all the results from the first, we can develop theories to
explain the emission from astrophysical objects such as stars, planetary
nebulae or, as for this work, blazars.

The theory of emission processess that occur between interacting non-thermal
distributions of lepto-hadronic particles has evolved over many years
(see e.g. \citealp{jones_calculated_1968,lightman_pair_1987,mastichiadis_x-ray_1992}),
as have methods to solve these systems numerically (e.g. \citealp{mastichiadis_self-consistent_1995}).
More recently, the increase in computational power has allowed for
the development of more complete models with better approximations
to the different processes and models that deal explicitly with more
kinds of particles, such as \citet{mastichiadis_spectral_2005,dimitrakoudis_time-dependent_2012,cerruti_hadronic_2015,petropoulou_photohadronic_2015}
and \citet{gao_direct_2017}.

In most cases, the computation of a final steady state for a model
given a series of parameters is time-consuming, which has prevented
the use of complex statistical tools in order to analyze the data,
the models, the parameters and their relations. Although some studies
use grid-based searches \citet{gao_modelling_2019} or genetic algorithms
\citet{rodrigues_blazars_2019} in order to systematically look for
the set of parameters that best fit the data, numerous studies still
fit models by eye \citet{keivani_multimessenger_2018,petropoulou_comprehensive_2020}.

In this work we introduce a novel code to model the behaviour of different
particle species in a relativistic plasma, \texttt{Katu}\footnote{Basque word for 'cat' https://en.wiktionary.org/wiki/katu\#Basque}.
\texttt{Katu} aims to be an efficient, highly-configurable model that
simulates the kinetic equations in a time-dependent manner.

Besides \texttt{Katu}, we also introduce two wrappers over commonly
used bayesian statistical packages: \texttt{multinest} \citet{feroz_importance_2013}
and \texttt{emcee} \citet{foreman-mackey_emcee:_2013}. This allows
us to not only have a powerful tool for simulating the emission from
plasmas, but to have a full suite ready to quickly analyse and fit
datasets producing high level statistical information. By having comprehensive
statistical information that goes beyond assigning a likelihood value
to a set of parameters, we also obtain converged probability distributions
of the parameters and their correlations along with a measure of their
Bayesian evidence that can be used for model comparisons..

This paper is structured as follows. We start in \secref{Physics}
by presenting the physics of the model, in particular, the kinetic
equation. We continue by presenting the two numerical approaches that
we take for solving the kinetic equation in \secref{Numerics}, with
some of the details of the implementation in \secref{Computational_tricks}.
Some simple tests of the numerical schemes we implement are shown
in \secref{Tests}. Next, an overview of bayesian inference, along
with a simple presentation of \texttt{emcee} and \texttt{multinest}
is presented in \secref{Model-Fitting}. We continue with a demonstration
of the suite by analysing the data of Mrk 421 during 2009 in \secref{Demonstration}.
We wrap up with our conclusions in \secref{Conclusions}. Additionally,
we include a full explanation of all the processes that we solve in
Appendix \ref{sec:Detailed-Description} and the configuration options
for \texttt{Katu} in Appendix \ref{sec:Configuration-File}.

\section{Physics of the model}{\label{sec:Physics}}

The model that the code \texttt{Katu} simulates is based on solving
the kinetic equation associated with $16$ different species of particles:
photons, protons, neutrons, electrons positrons, charged and neutral
pions, muons and antimuons, separated between left and right handed\footnote{The decay rates from pions and to neutrinos are different depending
on the handedness of the muons}, and electron and muon neutrinos and antineutrinos. The general form
of a kinetic equation is:

\begin{equation}
\frac{\partial n}{\partial t}=\mathcal{Q}_{e}+\mathcal{Q}_{i}+\mathcal{L}-\frac{n}{\tau_{esc}}-\frac{n}{\tau_{dec}}-\frac{\partial}{\partial\gamma}\left(\frac{\partial\gamma}{\partial t}n\right).\label{eq:kinetic_equation}
\end{equation}

Here, $\mathcal{Q}_{e}$ is an external injection of particles; $\mathcal{Q}_{i}$
is an internal injection of particles, that is particles that are
produced from processes inside the simulated volume; $\mathcal{L}$
is the term that denotes the loss of particles due to internal processes;
$\tau_{esc}$ represents the escape timescale of the particle, which
we take to be a multiple of the free escape timescale $\left(\tau_{esc}=\sigma\tau_{free,esc}\right)$;
$\tau_{dec}$ represents the decay timescale of the particle; and
$\partial\gamma/\partial t$ represents continuous gains and losses
of energy that affect the particles, where $\gamma$ represents the
energy of the particle normalized to its energy at rest.

In our model, the last term is used to represent accelerations proportional
to the energy of the particle, and losses due to synchrotron emission
and inverse Compton scattering, which are quadratic in energy. Thus,
we take:
\begin{equation}
\frac{\partial\gamma}{\partial t}=\frac{1}{\tau_{acc}}\gamma+S\gamma^{2},\label{eq:energy_change}
\end{equation}
where $\tau_{acc}$ represents the acceleration timescale and $S$
represents the proportionality constant between synchrotron and inverse
Compton losses and the energy squared.

Usually, the expressions for photon injection and losses are given
in terms of emission and absorption coefficients $\left(j_{\nu}\text{ and }\alpha_{\nu}\right)$,
which can be related to the injection and losses terms as:

\begin{align}
\mathcal{Q}_{i} & \coloneqq\left|\frac{\partial n_{\gamma}\left(\varepsilon\right)}{\partial t}\right|_{inj}=\frac{4\pi}{\varepsilon m_{e}c^{2}}\frac{d\nu}{d\varepsilon}j_{\nu}\left(\nu\right)=\frac{2}{\hbar\varepsilon}j_{\nu}\left(\nu\right),\\
\mathcal{L} & \coloneqq\left|\frac{\partial n_{\gamma}\left(\varepsilon\right)}{\partial t}\right|_{abs}=c\alpha_{\nu}\left(\nu\right)n_{\gamma}\left(\varepsilon\right),\label{eq:L_for_photons}
\end{align}
where $\varepsilon$ is the energy of the photons normalized to the
energy of an electron at rest $\left(\varepsilon=E_{\nu}/m_{e}c^{2}\right)$
and $\hbar$ is the reduced Plank constant.

\bigskip{}

In general, models that simulate the emission from relativistic plasmas
can be separated in three types depending on their treatment of the
primary and secondary particles. With primary and secondary particles
we refer, respectively, to those that are set up at the beginning
of the simulation or are injected from outside versus those that are
created by any internal process.

Of the first type we have the static models, where the primary particles
are set to a static distribution and the resulting photon field is
derived. Typically, no secondary particles are included. Effectively,
this usually corresponds to fixing the populations of electrons and
deriving the resulting photon population by applying \eqref{kinetic_equation}
with $\partial n_{\gamma}/\partial t=0$. The main disadvantage of
this kind of models is, evidently, that the primary particles can
not evolve and phenomena such as dynamic cascades can only be included
in an approximate manner.

The second type uses a hybrid approach where a population of primary
particles is fixed but a population of secondaries is allowed to vary
dynamically allowing the production of cascades of particles. Some
examples of these models are \citet{dimitrakoudis_time-dependent_2012,cerruti_hadronic_2015}.

The final option when modelling the system is to allow both primary
and secondary particles to vary with time. Examples of this third
type of models can be found in \citet{mastichiadis_self-consistent_1995,gao_direct_2017}.

Aiming to be comprehensive, we have set up \texttt{Katu} to include
a large number of processes in a large amount of detail. Appendix
\ref{sec:Detailed-Description} provides an in-depth description of
each process, along with the equations that are implemented in \texttt{Katu}.
For general reference, we recommend \citet[chapter 3]{rybicki_radiative_1979,Bottcher_Book}
and \citet{ghisellini_radiative_2013}. \Tabref{List-of-processes}
presents all the processes that we simulate along with the main references
that we have consulted.

\begin{table}
\begin{centering}
\begin{tabular}{c>{\centering}p{4cm}}
\toprule 
Process & References\tabularnewline
\midrule
\midrule 
Synchrotron Emission  & \citet{crusius_synchrotron_1986}\tabularnewline
and Absorption & \citet{Bottcher_Book}\tabularnewline
\midrule 
\multirow{2}{*}{Inverse Compton Scattering} & \citet{jones_calculated_1968}\tabularnewline
\cmidrule{2-2} 
 & \citet{Bottcher_Book}\tabularnewline
\midrule 
Two Photon Pair Production & \citet{bottcher_pair_1997}\tabularnewline
\midrule 
Pair Annihilation & \citet{svensson_pair_1982}\tabularnewline
\midrule 
Photo-meson Production & \citet{hummer_simplified_2010}\tabularnewline
\midrule 
Bethe-Heitler Pair Production & \citet[Section IV]{kelner_energy_2008}\tabularnewline
\midrule 
\multirow{2}{*}{Pion \& Muon decays} & \citet{lipari_flavor_2007}\tabularnewline
\cmidrule{2-2} 
 & \citet{hummer_simplified_2010}\tabularnewline
\bottomrule
\end{tabular}
\par\end{centering}
\caption{\label{tab:List-of-processes}List of processes and References.}

\end{table}

\section{Numerics}{\label{sec:Numerics}}

\Eqref{kinetic_equation} has a very complex form which can not be
solved analytically. Therefore, we must solve it numerically. Depending
on the type of particle whose population we want to evolve, certain
terms will not appear and the equation will be simpler to solve. But
first, we show the general solution of \eqref{kinetic_equation} in
terms of exponential integrators and an alternative solution using
an implicit scheme.

First, we substitute \eqref{energy_change} in \eqref{kinetic_equation},
expand the derivative in energy and merge similar terms:
\begin{align}
\frac{\partial n}{\partial t} & =\mathcal{Q}+\mathcal{L}+An+B\frac{\partial n}{\partial\gamma},
\end{align}
where:
\begin{align}
A & =-\frac{1}{\tau_{esc}}-\frac{1}{\tau_{dec}}-\frac{1}{\tau_{acc}}-2S\gamma,\\
B & =-\frac{\gamma}{\tau_{acc}}-S\gamma^{2}.
\end{align}

In general, every term in this equation depends on time, whether explicitly
(for example, if external injection is varied over time), through
a variable of the model (for example, $\tau_{esc}$ depends on the
size of the volume we are simulating, which can evolve over time)
or through the dependency on the population of other particles.

The loss term $\mathcal{L}$ can be separated as $\mathcal{L}=Ln$
(see for example \eqref[s]{L_for_photons} and \ref{eq:IC_losses-1})
and $L$ can be grouped with $A$. As a final step in transforming
the original equation, we note that $B$ depends on $\gamma$, so
we can change the derivative to logarithmic space. Additionally, we
know that it is common that the dependency of the population on the
energy takes the form of a power law. In this case, it is best to
take the derivatives of the logarithm of the population with respect
to the logarithm of the energy as it tends to be a small constant
that does not depend upon the population or the energy. Then we have
the two compact forms of the kinetic equation:
\begin{align}
\frac{\partial n}{\partial t} & =\mathcal{Q}+A'n+B'\frac{\partial n}{\partial\ln\gamma}.\label{eq:kinetic_equation_compact}\\
\frac{\partial\ln n}{\partial t} & =\frac{\mathcal{Q}}{n}+A'+B'\frac{\partial\ln n}{\partial\ln\gamma}.\label{eq:kinetic_equation_compact_log_log}
\end{align}
where:
\begin{align}
A' & =L-\frac{1}{\tau_{esc}}-\frac{1}{\tau_{dec}}-\frac{1}{\tau_{acc}}-2S\gamma,\\
B' & =-\frac{1}{\tau_{acc}}-S\gamma.
\end{align}

In the first case (equation \ref{eq:kinetic_equation_compact}), by
taking the derivative as an operator, we have a simple equation that
can be solved with an exponential integrator. This solution, under
the assumptions of constant $\mathcal{Q}$, $A'$ and $B'$ is exact
(unless not binned numerically over a finite number over energies
with a given representation of the derivative). This solution is very
interesting because, for cases where there is no acceleration or cooling,
that is, there is no derivative term, this remains \emph{exact} when
implemented numerically. This is the case for neutral particles, which
have $\tau_{acc}=0$ and $S=0$, and thus $B'=0$.

Where this is not the case and we have to keep the derivative, even
though the solution that we find through the exponential integrator
is correct, its calculation involves obtaining the exponential and
the inverse of an operator, which is very costly in computational
time, so we explore other solutions in the form of an implicit scheme
of \eqref{kinetic_equation_compact_log_log}.

\subsection{Exponential Integrator Solutions}

Taking the derivative as an operator, the solution of \eqref{kinetic_equation_compact}
is:
\begin{align}
n\left(t_{0}+\Delta_{t}\right) & =\exp\left[\int_{t_{0}}^{t_{0}+\Delta_{t}}\left(A'+B'\frac{\partial}{\partial\ln\gamma}\right)dt'\right]n\left(t_{0}\right)+\nonumber \\
 & \int_{t_{0}}^{t_{0}+\Delta_{t}}\exp\left[\int_{t'}^{t_{0}+\Delta_{t}}\left(A'+B'\frac{\partial}{\partial\ln\gamma}\right)dt''\right]\mathcal{Q}dt'\label{eq:Exponential_Integrator_Solution}
\end{align}
where $\mathcal{Q}$, $A'$ and $B'$ all depend on time.

To obtain the first order solution to \eqref{Exponential_Integrator_Solution},
we consider that every term varies very little between $t_{0}$ and
$t_{0}+\Delta_{t}$ and can be taken as constant. Then we obtain:
\begin{align}
n\left(t_{0}+\Delta_{t}\right) & =\exp\left[\Delta_{t}\left(A'+B'\frac{\partial}{\partial\ln\gamma}\right)\right]n\left(t_{0}\right)+\nonumber \\
 & \left\{ \exp\left[\Delta_{t}\left(A'+B'\frac{\partial}{\partial\ln\gamma}\right)\right]-I\right\} \left[A'+B'\frac{\partial}{\partial\ln\gamma}\right]^{-1}\mathcal{Q},
\end{align}
where the exponential and the inverse of an operator are defined through
their Taylor expansions:
\begin{align}
\exp\left[\Delta_{t}\left(A'+B'\frac{\partial}{\partial\ln\gamma}\right)\right] & \coloneqq\sum_{i=0}^{\infty}\frac{1}{i!}\left[\Delta_{t}\left(A'+B'\frac{\partial}{\partial\ln\gamma}\right)\right]^{i},\\
\left(A'+B'\frac{\partial}{\partial\ln\gamma}\right)^{-1} & \coloneqq\sum_{i=0}^{\infty}\left(I-A'-B'\frac{\partial}{\partial\ln\gamma}\right)^{i}.
\end{align}

Note that both $A'$ and $B'$ depend on $\gamma$, so the cross terms
of $B'\frac{\partial}{\partial\ln\gamma}A'$ and $B'\frac{\partial}{\partial\ln\gamma}\left(B'\frac{\partial}{\partial\ln\gamma}\right)$
are not null.

This step usually involves the exponentiation and inversion of an
operator, which is very costly computationally. But, as we have said
in the previous section, this solution is very interesting in the
case that $B'=0$ as it reduces to:
\begin{equation}
n\left(t_{0}+\Delta_{t}\right)=\exp\left[\Delta_{t}A'\right]n\left(t_{0}\right)+\frac{\exp\left[\Delta_{t}A'\right]-I}{A'}\mathcal{Q},
\end{equation}
where the exponential and the inverse are easily calculable. This
is the solution that we employ for neutral particles (neutrons, neutral
pions and neutrinos).

\subsection{Implicit Method Solution}

An alternative way of solving \eqref{kinetic_equation} is to discretize
the derivative in energy as the first step and then look for a solution.
In order to make the solution stable for high values of $\Delta_{t}$,
we work from an implicit expression for the derivative. In this case,
it is better to work with logarithmic derivatives, as this makes the
numerical solution more stable. To account for the possible flows
of particles due to the acceleration and cooling, we have to look
for an equilibrium point where the two fluxes converge and separate
the solution in two parts at the equilibrium energy.

We note from \eqref{energy_change} that there is an equilibrium point
in energy at $\gamma_{eq}=-\frac{1}{\tau_{acc}S}$ (remember that
$S<0$) and that for $\gamma<\gamma_{eq}$, we have that $\frac{\partial\gamma}{\partial t}>0$
and conversely, for $\gamma>\gamma_{eq}$, we have that $\frac{\partial\gamma}{\partial t}<0$.
This gives us the energy direction and thus, the direction in which
we have to take the derivatives when discretizing. This is equivalent
to upwind schemes for flows of matter, although in energy space.

In this case the discretized version of \eqref{kinetic_equation_compact_log_log},
taking index $i$ as the index of the highest energy that is lower
than $\gamma_{eq}$, now reads:
\begin{equation}
\frac{\partial\ln n_{j}^{t}}{\partial t}=\begin{cases}
\frac{\mathcal{Q}_{j}}{n_{j}^{t+1}}+A_{j}'+B_{j}'\left(\frac{\ln n_{j}^{t+1}-\ln n_{j-1}^{t+1}}{\Delta\ln\gamma}\right) & \text{for }j<i\\
\frac{\mathcal{Q}_{j}}{n_{j}^{t+1}}+A_{j}'+B_{j}'\left(\frac{\ln n_{j+1}^{t+1}-\ln n_{j}^{t+1}}{\Delta\ln\gamma}\right) & \text{for }j>i
\end{cases}.
\end{equation}

Discretizing now in time, and isolating the terms in $\ln n_{j}^{t+1}$
we obtain:
\begin{align}
\ln n_{j}^{t+1} & =\begin{cases}
\ln n_{j}^{t}+\Delta_{t}\left[A_{j}'-B_{j}'\frac{\ln n_{j-1}^{t+1}}{\Delta\ln\gamma}\right]+\\
\phantom{\ln n_{j}^{t}+\;\,}\Delta_{t}\left[\frac{\mathcal{Q}_{j}}{n_{j}^{t+1}}+B_{j}'\frac{\ln n_{j}^{t+1}}{\Delta\ln\gamma}\right] & \text{for }j<i\\
\ln n_{j}^{t}+\Delta_{t}\left[A_{j}'+B_{j}'\frac{\ln n_{j+1}^{t+1}}{\Delta\ln\gamma}\right]+\\
\phantom{\ln n_{j}^{t}+\;\,}\Delta_{t}\left[\frac{\mathcal{Q}_{j}}{n_{j}^{t+1}}-B_{j}'\frac{\ln n_{j}^{t+1}}{\Delta\ln\gamma}\right] & \text{for }j>i
\end{cases}.
\end{align}

For convenience, we define the following constants:
\begin{align}
C_{\pm} & \coloneqq A_{j}'\pm B_{j}'\frac{\ln n_{j\pm1}^{t+1}}{\Delta\ln\gamma},\\
D_{\pm} & \coloneqq\pm\frac{\Delta_{t}}{\Delta\ln\gamma}B_{j}'-1.
\end{align}

The solution can now be found in terms of the Lambert W-function:
\begin{equation}
n_{j}^{t+1}=\begin{cases}
-\frac{\Delta_{t}\mathcal{Q}_{j}}{D_{+}W\left[-\frac{\Delta_{t}\mathcal{Q}_{j}}{D_{+}}\exp\left(\frac{\ln n_{j}^{t}+\Delta_{t}C_{-}}{D_{+}}\right)\right]} & \text{for }j<i\\
-\frac{\Delta_{t}\mathcal{Q}_{j}}{D_{-}W\left[-\frac{\Delta_{t}\mathcal{Q}_{j}}{D_{-}}\exp\left(\frac{\ln n_{j}^{t}+\Delta_{t}C_{+}}{D_{-}}\right)\right]} & \text{for }j>i
\end{cases}.
\end{equation}
The Lambert function has the property that $\ln W\left(x\right)=\ln x-W\left(x\right)$
for $x>0$, so we can obtain the logarithm of the population as:
\begin{align}
\ln n_{j}^{t+1} & =\begin{cases}
-\frac{\ln n_{j}^{t}+\Delta_{t}C_{-}}{D_{+}}+\\
\quad W\left[-\frac{\Delta_{t}\mathcal{Q}_{j}}{D_{+}}\exp\left(\frac{\ln n_{j}^{t}+\Delta_{t}C_{-}}{D_{+}}\right)\right] & \text{for }j<i\\
-\frac{\ln n_{j}^{t}+\Delta_{t}C_{+}}{D_{-}}+\\
\quad W\left[-\frac{\Delta_{t}\mathcal{Q}_{j}}{D_{-}}\exp\left(\frac{\ln n_{j}^{t}+\Delta_{t}C_{+}}{D_{-}}\right)\right] & \text{for }j>i
\end{cases}.
\end{align}

By imposing some boundary conditions for $\ln n_{0}^{t+1}$ and $\ln n_{n}^{t+1}$,
the rest of the populations can be solved for in a straightforward
manner. This is the solution that we employ for charged particles
(protons, electrons and positrons, muons and charged pions).

\section{Computational Implementation}{\label{sec:Computational_tricks}}

\subsection{Initialization}

\texttt{Katu} is initialized according to a configuration file with
an extensive range of options (explained in Appendix \ref{sec:Configuration-File}).
Here, we explain how the configuration settings are translated into
the model variables.

The initial electron and proton populations are set using background
density $\left(\rho\right)$, ratio of protons to electrons $\left(\eta\right)$
and the shapes of their distributions over energies as set in the
configuration file. The resulting initial number densities for electrons
and protons respectively are therefore given by:
\begin{align}
n_{e}\left(\gamma_{e}\right)= & N_{e}f_{e}\left(\gamma_{e}\right)=\frac{\rho}{m_{e}+\eta m_{p}}f_{e}\left(\gamma_{e}\right),\\
n_{p}\left(\gamma_{p}\right)= & N_{p}f_{p}\left(\gamma_{p}\right)=\frac{\eta\rho}{m_{e}+\eta m_{p}}f_{p}\left(\gamma_{p}\right).
\end{align}
Here $f_{e}$ is the normalized distribution function for electrons
and $N_{e}$ the total electron number density (likewise for protons).

\bigskip{}

For the external injection of particles, we follow a similar procedure,
except that instead of using the density of the fluid $\rho$, we
use the total particle luminosity $L$. Then, if we inject protons
and electrons in a plasma volume $\left(V\right)$ we have:
\begin{align}
\frac{L}{V} & =m_{e}c^{2}\int_{\gamma_{e,min}}^{\gamma_{e,max}}\gamma_{e}Q_{e,0}f_{e}\left(\gamma_{e}\right)d\gamma_{e}+\nonumber \\
 & \phantom{=\;}m_{p}c^{2}\int_{\gamma_{p,min}}^{\gamma_{p,max}}\gamma_{p}Q_{p,0}f_{p}\left(\gamma_{p}\right)d\gamma_{p},
\end{align}
where $Q_{0,e}$ and $Q_{0,p}$ are respectively the number densities
of injected electrons and protons, respectively. Injecting $\eta$
protons per electron $\left(Q_{p,0}=\eta Q_{e,0}\right)$, and calculating
the integrals, we have that:
\[
\frac{L}{V}=Q_{e,0}\left\langle \gamma_{e}\right\rangle m_{e}c^{2}+\eta Q_{e,0}\left\langle \gamma_{p}\right\rangle m_{p}c^{2},
\]
where $\left\langle \gamma\right\rangle $ is the average energy of
the injected particles. Solving for $Q_{e,0}$, we can obtain that
the injection distributions are given by:
\begin{align}
Q_{e}\left(\gamma_{e}\right) & =\phantom{\eta}Q_{e,0}f_{e}\left(\gamma_{e}\right),\\
Q_{p}\left(\gamma_{p}\right) & =\eta Q_{e,0}f_{p}\left(\gamma_{p}\right).
\end{align}

\bigskip{}

The external injection of photons is much simpler as only one particle
kind is involved and thus we have:
\begin{equation}
Q_{\gamma}\left(\varepsilon\right)=\frac{L_{\gamma}}{\left\langle \varepsilon\right\rangle Vm_{e}c^{2}},
\end{equation}
where $L_{\gamma}$ is the luminosity of the external photon field
and $\left\langle \varepsilon\right\rangle $ is its normalized average
energy.

\bigskip{}

The value of the free escape timescale is calculated from the volume
as:
\begin{equation}
t_{free,esc}=\begin{cases}
\frac{3}{4}\frac{R}{c} & \text{if sphere}\\
\frac{\pi}{4}\frac{h}{c} & \text{if disk}
\end{cases},
\end{equation}
where $R$ is the radius of the sphere and $h$ is the height of the
disk.

\subsubsection{Numerical Initialization}

Even though mathematically the values that we have found for the different
injections, initial number densities and so on are exact, some of
the distributions that we use do not have formulas with which to get
their normalized form. This means that we actually have to numerically
calculate them.

Then, to obtain the values of the injection in the population distribution,
we bin the corresponding injection onto the population binning.

\bigskip{}

Only the boundary values for the ranges of the proton, electron and
photon populations are set by the user (see Appendix \ref{sec:Configuration-File}).
For other particles, their boundaries have to be calculated so that
we are sure that the injection of secondary particles does not fall
outside the range of their distributions. In particular we take:
\begin{align}
\gamma_{n,min} & =\gamma_{p,min},\\
\gamma_{n,max} & =\gamma_{p,max},\\
\gamma_{e^{+},min} & =\gamma_{e^{-},min},\\
\gamma_{e^{+},max} & =\gamma_{e^{-},max},\\
\gamma_{\pi,min} & =\max\left(0.2\cdot\gamma_{p,min}\cdot\frac{m_{p}}{m_{\pi}},10^{0.5}\right),\\
\gamma_{\pi,max} & =0.6\cdot\gamma_{p,max}\cdot\frac{m_{p}}{m_{\pi}},\\
\gamma_{\mu,min} & =\max\left(\gamma_{\pi,min}\cdot\frac{m_{\pi}}{m_{\mu}},10^{0.5}\right),\\
\gamma_{\mu,max} & =\gamma_{\pi,max}\cdot\frac{m_{\pi}}{m_{\mu}},\\
\gamma_{\nu,min} & =\min\left(\gamma_{\pi,min}\cdot\frac{m_{\pi}}{m_{e}},\gamma_{\mu,min}\cdot\frac{m_{\mu}}{m_{e}}\right),\\
\gamma_{\nu,max} & =\max\left(\gamma_{\pi,max}\cdot\frac{m_{\pi}}{m_{e}}\cdot\left(1-\frac{m_{\mu}^{2}}{m_{e}^{2}}\right),\gamma_{\mu,max}\cdot\frac{m_{\mu}}{m_{e}}\right),
\end{align}
where $\gamma_{n},\gamma_{e^{+}},\gamma_{\pi},\gamma_{\mu}$ and $\gamma_{\nu}$
refer to the limits for neutrons, positrons, pions, muons and neutrinos
respectively. It must be noted that we do not modify the higher limit
for the electron population, and thus, the user must make sure that
the electrons created by the Bethe-Heitler process do not fall outside
the boundaries.

\bigskip{}

In order for the numerical schemes to be stable, the maximum possible
timestep has to be selected as lower than the free escape timescale:
\begin{equation}
\Delta_{t,max}=\max\left(\Delta_{t,conf},t_{free,esc}\right).
\end{equation}

\subsection{Stopping Criteria}

\texttt{Katu} evolves the population of the particles according to
the kinetic equation (\ref{eq:kinetic_equation}). As such, it would
continuing stepping indefinitely if no stopping criteria are provided.
In our modelling, and in the examples that we provide with the code,
we have added two stopping criteria.

In the first case, we stop the modelling if the elapsed time is bigger
than a $t_{max}$ set in the configuration file.

In the second case, we stop if the simulation reaches a steady state
with a tolerance better than the one required in the configuration
file. For this, we check the evolution of three species of particles:
electrons, photons and neutrinos, according to the following formula:
\begin{equation}
\sqrt{\int\left[\frac{1}{n\left(x\right)}\frac{\partial n\left(x\right)}{\partial t}\right]^{2}dx}<\text{tol},
\end{equation}
where $n$ is the number density for a given type of particle.

\subsection{Computational Optimizations }{\label{subsec:Computational-Tricks}}

A number of strategies have been implemented to optimize the computational
efficiency of our code.

\subsubsection{Look-Up-Tables}

Most, if not all, processes have an equation that fall into one of
the following forms:
\begin{align}
\frac{\partial n\left(x\right)}{\partial t} & =\int_{y_{min}}^{y_{max}}n\left(y\right)R\left(x,y\right)\,dy,\label{eq:lut_1}\\
\frac{\partial n\left(x\right)}{\partial t} & =\int_{y_{min}}^{y_{max}}n\left(y\right)\int_{z_{min}}^{z_{max}}n\left(z\right)R\left(x,y,z\right)\,dzdy,\label{eq:lut_2}
\end{align}
where $R$ is a generic function describing an interaction between
particles.

For example, synchrotron emission and absorption and almost all loss
terms follow the first form, whereas all the processes that require
two particles interacting follow the second form. The important part
is that the function $R$ does only depend on the energy of the implied
particles and thus, can be tabulated as part of the start-up of the
modelling process once and be used repeatedly.

For the more extreme cases of complicated functions, such as for the
Bethe-Heitler process, we create a super-table that encompasses the
most likely boundaries for the energy and interpolate it onto the
energy grid.

Reducing every processes to the form of \eqref[s]{lut_1} and (\ref{eq:lut_2})
allows us to transform their implementations into a loop over the
produced particles along with two nested integrals. In practice, this
means that the worst case scenario for a process is to scale cubically
with the number of bins for a particle population. The processes currently
included in Katu scale at most quadratically with the population sizes
of photons and electrons and linearly with the other particles.

\subsubsection{Threading}

During a single time step (assumed to be sufficiently small) all processes
are treated as independent and can therefore be computed in parallel.
As part of the start-up process, we create a series of workers and
add them to a pool to wait for work. When the moment of calculating
each process arrives, the manager thread adds each process to a queue
and wakes up the workers, which take a work item from the queue and
complete it. Then, the workers check the queue until it is empty and
go back to sleep, with the last one signalling the manager thread
that it can all have finished and the program can continue.

\section{Tests of the model}{\label{sec:Tests}}

The model on which we base our code included linear acceleration,
handled through an acceleration timescale $t_{acc}$, and cooling
caused by the synchrotron and inverse Compton processes. These terms,
while mathematically simple, pose a computational challenge to calculate
correctly, as discussed in \secref{Numerics}. Hence, we concentrate
our tests on verifying acceleration and cooling.

All tests have an extremely low value of background plasma $\left(\rho=10^{-38}\right)$,
injection $\left(L=10^{30}\right)$ and ratio of protons to electrons
$\left(\eta=10^{-10}\right)$. This is in order to have a low population
of pure electrons and a minimal quantity of photons, so that the losses
through Inverse Compton cooling are minimal and the cooling behaviour
can be controlled by setting the magnetic field strength. The geometry
of the blob is chosen as a sphere with $R=10^{16}$, which gives $t_{esc}\approx250173$
s.

\bigskip{}

To test the acceleration code, we inject electrons at low energy $\gamma_{e,inj,max}=10^{2}$
and wait until the system has reached a steady state. According to
\citet{mastichiadis_self-consistent_1995}, above the maximum energy
of injection, the electron population should adopt a power law behaviour
with index $1+t_{acc}/t_{esc}$.

\begin{figure}
\begin{centering}
\includegraphics[width=1\columnwidth]{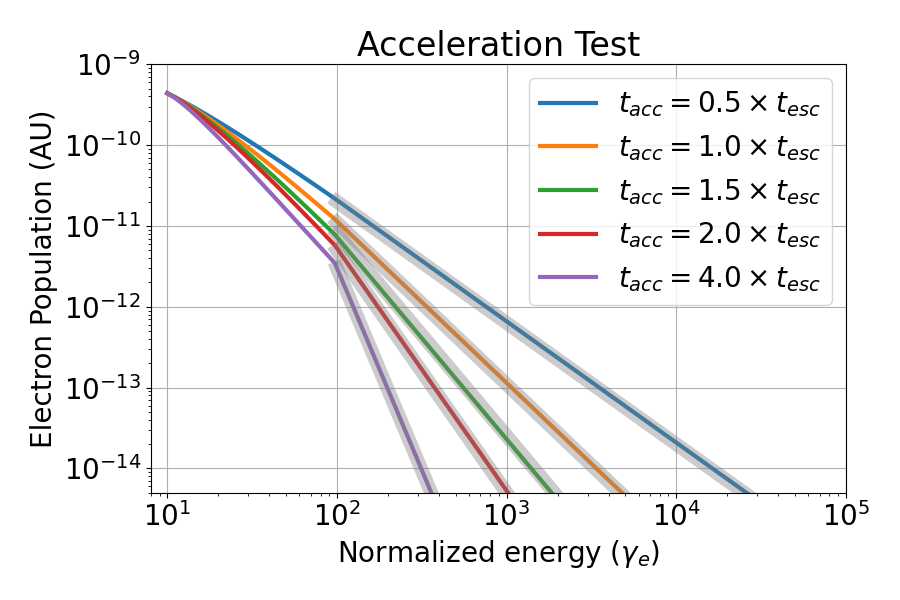}
\par\end{centering}
\caption{\label{fig:Acceleration_Tests}Acceleration tests. Solid lines correspond
to the output of our tests. The bands represent power laws with index
$1+t_{acc}/t_{esc}$.}

\end{figure}

\Figref{Acceleration_Tests} shows the result of the tests with different
ratios of $t_{acc}/t_{esc}$ implemented by varying the value of $t_{acc}$.
We obtain a near-exact match between the numerical curves and their
theoretical values.

\bigskip{}

To test the cooling code, we take a similar approach to the acceleration
test, but we inject electrons only at high energies $\left(\gamma_{e,inj,min}=10^{4}\right)$
with a power law index of $p=2.3$. The first part of this test studies
the behaviour of the particle population below the injection point.
By assuming a simplified version of \eqref{kinetic_equation} in which
there is neither injection nor acceleration and the only losses are
due to escape and cooling, we can obtain a steady state solution for
the population as:
\begin{equation}
n\left(\gamma\right)=C\exp\left(\frac{1}{S\tau_{esc}\gamma}\right)\frac{1}{\gamma^{2}}\label{eq:population_low_energy_steady_state}
\end{equation}
where $C$ is a constant. In the absence of other contributions to
the particle population, this solution fully represents the particle
population distribution below the lower cut-off of the injection.
For values of $S\tau_{esc}\gamma\gg1$, the exponential term tends
to $1$ and we obtain the result that the population behaves like
a power law with index $2$. For other cases, the exponential term
is important and causes an exponential drop in the population towards
lower energies. Physically, the first case represents that the cooling
timescale is lower than the escape timescale where particles cool
faster than they escape, which gives the power law behaviour. In the
second case, we have that particles can escape much faster than they
cool, so the population drops exponentially towards lower energies.

\begin{figure}
\begin{centering}
\includegraphics[width=1\columnwidth]{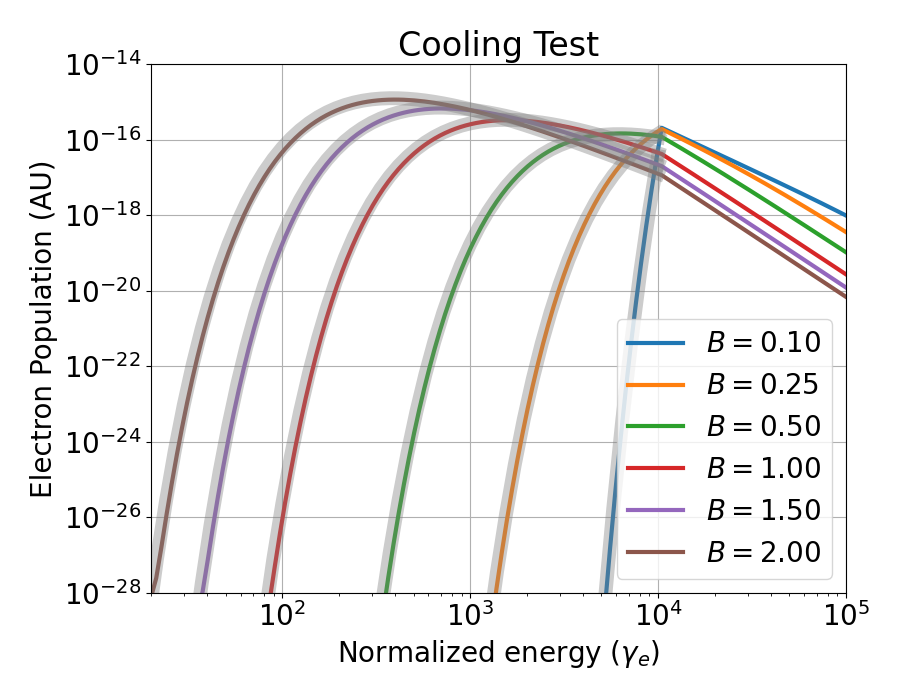}
\par\end{centering}
\caption{\label{fig:Cooling_Tests_low_energy}Cooling tests below the injection
lower limit. Solid lines correspond to the output of our tests. The
bands represent \eqref{population_low_energy_steady_state}.}
\end{figure}

In \figref{Cooling_Tests_low_energy} we show the resulting population
of electrons for six different values of the magnetic field where
on one extreme $\left(B=0.1\right)$ we have that the main driver
of the particle population is escape, whereas for the other $\left(B=2\right)$
it is cooling. The bands follow \eqref{population_low_energy_steady_state}
with $C$ chosen as to match approximately the value of the population
at the lower end of the injection point. As we can see, the match
is quite good, which demonstrates the validity of our approach for
the cooling algorithms.\bigskip{}

The second part of the cooling test involves the behaviour of the
particles in the injection range. For this range, we have again two
behaviours depending on the cooling and escape timescales. The transition
between these two behaviours is defined by a cooling break energy,
which can be estimated as $\gamma_{c}\approx1/\left(1-p_{inj}\right)S\tau_{esc}$.
Hence, keeping $p_{inj}$ and $\tau_{esc}$ constant, we can vary
the value of the magnetic field to control the position of the cooling
break. Below the cooling break, we have a population that is escape
dominated, and develops into a population that keeps the original
behaviour of the injection profile. Above the cooling break, the population
is cooling dominated and as a result the population follows a power
law steepened by $1$.\textcolor{red}{}

\begin{figure}
\begin{centering}
\includegraphics[width=1\columnwidth]{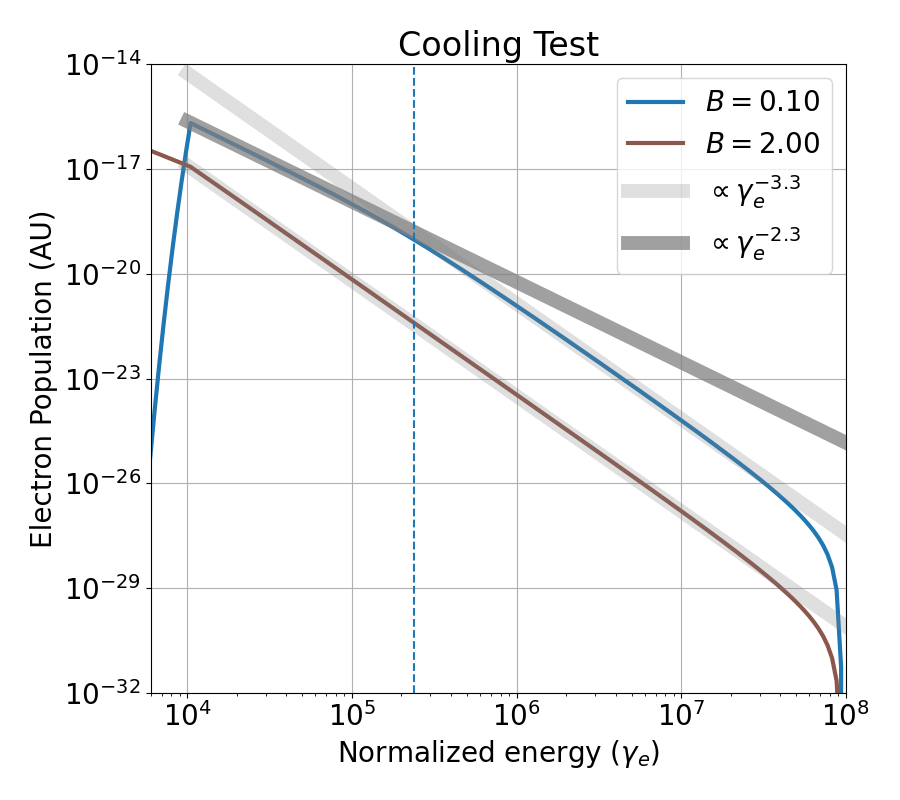}
\par\end{centering}
\caption{\label{fig:Cooling_Tests}Cooling tests in the injection range. Solid
lines correspond to the output of our tests. The bands represent the
theoretical approximations for the population. The vertical dashed
blue line corresponds to the cooling break for $B=0.1$.}
\end{figure}

For \figref{Cooling_Tests} we have tested two values of the magnetic
field $\left(B=0.1\text{ and }B=2\right)$. In the first case (blue
line), the cooling break is between the injection ends, so we get
two behaviours as expected. For the lower energy part, the resulting
population has an index of $2.3$ and for the higher energy part,
the index is $3.3$, just as expected. Below the injection lower limit,
the population of electrons escapes freely and that is reflected by
the abrupt drop in population. For the second case (brown line), the
cooling timescale is much lower and the cooling break is below the
lower injection limit. Hence, between the injection ends, we see that
the whole population has a index of $3.3$.

\bigskip{}

In addition to the above tests, which are designed to stress the acceleration
and cooling parts of our numerical schemes, we have also performed
tests aimed towards checking the correctness of the implementation
of the physical processes. In this case, we have checked that the
tests of \citet{dimitrakoudis_time-dependent_2012} pass. Further
demonstrations of the code can be found in \citet{jimenez-fernandez_bayesian_2020}.

\section{Bayesian Inference, Markov Chain Monte Carlo and Nested Sampling
analysis}{\label{sec:Model-Fitting}}

Bayesian inference provides a useful approach to parameter estimation
and model selection. In this section we describe some of the key concepts
of Bayesian inference and how it can be used in the task of fitting
a model. For the case of blazar modelling, we additionally describe
two different approaches that \texttt{Katu} is able to work with via
two available wrappers: Markov Chain Monte Carlo (MCMC) analysis in
the form of an ensemble sampler with affine invariance \citet{goodman_ensemble_2010}
and Nested Sampling (NS) analysis in the form of nested elliptical
sampling \citet{feroz_multimodal_2008}.

\subsection{Bayesian Inference}{\label{subsec:Nomenclature}}

We define a dataset $\mathcal{D}$ as the combination of observations
$y$ and their associated errors $\sigma$. A model $\mathcal{M}$
is defined as some function that can be applied to a set of parameters
$\theta$ to obtain some model predictions $y_{\mathcal{M}}$ that
can be compared with the dataset. Then, we can define the probability
that a dataset has been obtained given a model and a set or parameters
as:
\begin{equation}
P\left(\mathcal{D}|\theta,\mathcal{M}\right)\equiv\mathcal{L},
\end{equation}
where $\mathcal{L}$ is called the likelihood. Likewise, we can also
define the a priori probability, or belief, of the parameters that
we have used:
\begin{equation}
P\left(\theta|\mathcal{M}\right)\equiv\pi,
\end{equation}
where $\pi$ is denoted the prior. For example, when comparing leptonic
and lepto-hadronic models, the parameter that most determines this
division is $\eta$. If we have any a priori information suggesting
a certain distribution for $\eta$, we can include this information
as a prior. If not, the prior is typically set to be flat. Note however
that this does not mean that no prior assumptions were included: a
flat prior in logarithmic space will for example have a different
impact than a flat prior in linear space.

By integrating over the whole parameter space $\Theta$, we can obtain
the probability of obtaining a dataset given a model:
\begin{equation}
P\left(\mathcal{D}|\mathcal{M}\right)\equiv\mathcal{Z}=\int_{\Theta}P\left(\mathcal{D}|\theta,\mathcal{M}\right)P\left(\theta|\mathcal{M}\right)d\theta,\label{eq:evidence_expanded}
\end{equation}
where $\mathcal{Z}$ is the evidence of the model, or, simplifying
the notation:
\begin{equation}
\mathcal{Z}=\int_{\Theta}\mathcal{L}\left(\theta\right)\pi\left(\theta\right)d\theta.\label{eq:Evidence}
\end{equation}

The evidence can be understood as a weighted average of the likelihood
$\mathcal{L}$ over the whole parameter space with the prior $\pi$
as a weight function. Thus, only the regions where the product of
both the likelihood and the prior is high will mainly contribute to
the value of the evidence. By integrating over the whole parameter
space with a normalized prior, we are implicitly dividing by the volume
of the space so models with more parameters that do not improve the
likelihood will obtain a lower evidence.

By applying Bayes' theorem, we can find the probability of a model
given a dataset:
\begin{equation}
P\left(\mathcal{M}|\mathcal{D}\right)=\frac{P\left(\mathcal{D}|\mathcal{M}\right)P\left(\mathcal{M}\right)}{P\left(\mathcal{D}\right)},
\end{equation}
where $P\left(\mathcal{M}\right)$ and $P\left(\mathcal{D}\right)$
are the a priori probability, or belief, of the model and the data
respectively. Here, $P\left(\mathcal{D}\right)$ denotes the probability
of measuring the dataset $\mathcal{D}$ from the underlying physics
and the observation method. $P\left(\mathcal{M}\right)$ encapsulates
our prior belief in the model $\mathcal{M}$ that is not already reflected
in the priors of the parameters. For example, if we know that blazars
could be fit from two different models with a ratio of $1$ to $4$,
we can incorporate this information in this term.

Denoting two models by $a$ and $b$, we can take the ratio of their
probabilities to obtain:
\begin{equation}
R=\frac{P\left(\mathcal{M}_{a}|\mathcal{D}\right)}{P\left(\mathcal{M}_{b}|\mathcal{D}\right)}=\frac{\mathcal{Z}_{a}P\left(\mathcal{M}_{a}\right)}{\mathcal{Z}_{b}P\left(\mathcal{M}_{b}\right)}.
\end{equation}
Assuming that both models have the same a priori probability, we are
left with the ratio of evidences as a way of comparing two models,
so that if $R>1$ model $a$ is more likely and vice versa, if $R<1$,
model $b$ is more likely.

As with all criteria for model selection, the actual amount of (relative)
confidence in a given model that is linked to a given numerical value
of R remains subjective. Taking as a guideline \citet{bayes_factors},
we give the following references for the value of $R$: $1-3.2$,
`not worth more than a bare mention'; $3.2-10$, `substantial';
$10-100$, `strong' and $>100$, `decisive'.

\bigskip{}

By inverting \eqref{evidence_expanded} and the conditioning, we can
find the probability of a point in parameter space given a dataset
and a model:
\begin{equation}
P\left(\theta|\mathcal{D},\mathcal{M}\right)=\frac{P\left(\mathcal{D}|\theta,\mathcal{M}\right)P\left(\theta|\mathcal{M}\right)}{P\left(\mathcal{D}|\mathcal{M}\right)},
\end{equation}
which we call the posterior distribution of $\theta$: $\mathcal{P}\left(\theta\right)$.
Simplifying the notation, we obtain:
\begin{equation}
\mathcal{P}\left(\theta\right)=\frac{\mathcal{L}\left(\theta\right)\pi\left(\theta\right)}{\mathcal{Z}}.
\end{equation}

For the purpose of parameter estimation, it is not necessary to compute
the evidence. \bigskip{}

Assuming that the data follow a Gaussian distribution, we can define
the log-likelihood\footnote{The reader should not confuse the Likelihood $\left(\mathcal{L}\right)$
with the log-likelihood $\left(L\right)$.} associated between a dataset and the model predictions as:
\begin{equation}
L=-\frac{1}{2}\sum_{i}^{n}\left[\ln\left(2\pi\sigma_{i}^{2}\right)+\left(\frac{y_{i}-y_{\mathcal{M},i}}{\sigma_{i}}\right)^{2}\right].
\end{equation}
The first term is a constant for every log-likelihood, as it does
not depend on the model, only on the errors of the data. It can be
proven that this term appears as a constant multiplicative factor
upon the calculation of the evidence, so it divides out when comparing
the evidence of two models.

\subsection{MCMC and Nested Sampling}{\label{subsec:MCMC_emcee}}

MCMC and Nested Sampling are two different approaches to analyse,
using bayesian inference, a model given its likelihood. MCMC can sample
from a given function, which in this case it is the likelihood of
the model, and thus, we can reconstruct the shape of the likelihood
and the posterior probabilities of any of its parameters. Nested sampling
works by sampling from the likelihood function in a monotonically
increasingly way. By assigning a statistical weight to each of the
samples, the shape of the likelihood can be reconstructed and the
evidence of the model can be calculated.

\bigskip{}

Even though sampling using MCMC is simple in principle, it has some
drawbacks when the shape of the sampled distribution has various maxima
and/or is very different from a simple gaussian. In order to avoid
partially these problems, \citet{goodman_ensemble_2010} introduce
the 'Ensemble Samplers with Affine Invariance' and \citet{foreman-mackey_emcee:_2013}
creates, using ideas from that work, \texttt{emcee}. The MCMC steps
introduced in \citet{goodman_ensemble_2010} and implemented in \texttt{emcee}
allows for sampling efficiently from any distribution that can be
reduced to a multivariate gaussian by an affine transformation, increasing
greatly the distributions for which MCMC is efficient at the cost
of increasing the memory requirements. In particular, whereas simple
MCMC uses a 'walker' to traverse the parameter space, \texttt{emcee}
uses an ensemble of them.

Even though the affine invariant ensemble samplers and \texttt{emcee}
are a great statistical tool, they still have some problems inherited
from the basic MCMC analysis. First, MCMC tends to perform very badly
in situations where multiple modes exist, as it can take a long time
for a walker to go from one maximum to another because in its path
there is a volume of likelihood that is taken with lower probability
than other steps. Second, MCMC needs some period of burn-in until
the sampler starts to produce samples that are independent and representative
of the sampled probability distribution. Third, not every distribution
is well represented by affine transformations of a multivariate gaussian,
so there are cases where \texttt{emcee} will still perform not optimally.

Nevertheless, after some iterations, we obtain a chain of parameters
for each walker from which we can derive the sought for posterior
probabilities by simply plotting the appropiate histograms.

\bigskip{}

In contrast to MCMC analysis, nested sampled is aimed towards the
calculation of the evidence. The basic idea is that, by creating a
sequence of points with increasingly likelihood and by assigning a
weight to each of them depending on their position in the sequence,
the evidence can be calculated. As the value of the likelihood increases
in the sequence, the probability of drawing a new point with a better
value decreases. \texttt{multinest} avoids this problem by using a
set of points as 'live points' and using them to define ellipses from
where to drawn the new points \citep{feroz_multinest:_2009}. This
way, the sampling is much more efficient and the sequence can grow
faster.

As the set of 'live points' define the sampled volume, it can happen
that this volume can be separated into isolated sub-volumes. In these
cases, each of the sub-volumes is associated with one maximum of the
likelihood function, and Nested Sampling, and \texttt{multinest} by
extension, can detect and handle these separated maxima with ease.

Even though the length of the sequence can be made as big as wanted,
there is a point where the last points contribute very little to the
calculation of the evidence. \texttt{multinest}, by default, is set
to stop the calculation of the evidence when the best point of the
sequence increases the value of the log evidence by less than some
tolerance.

Each of the points in the sequence is assigned a weight, and thus,
can be used to create the histograms of the posterior probabilities
of the parameters.

\section{Application to Mrk 421}{\label{sec:Demonstration}}

\subsection{Dataset}{\label{subsec:Dataset}}

The blazar Markarian 421 (Mrk 421) has been observed for decades.
Mrk 421 is one of the nearest blazars to earth, and therefore one
of the brightest. Mrk 421 is a BL Lacertae object, which tend to be
variable in their flux levels and can typically be modelled without
a need for an external source of photons.

\begin{figure*}
\begin{centering}
\includegraphics[width=1\textwidth]{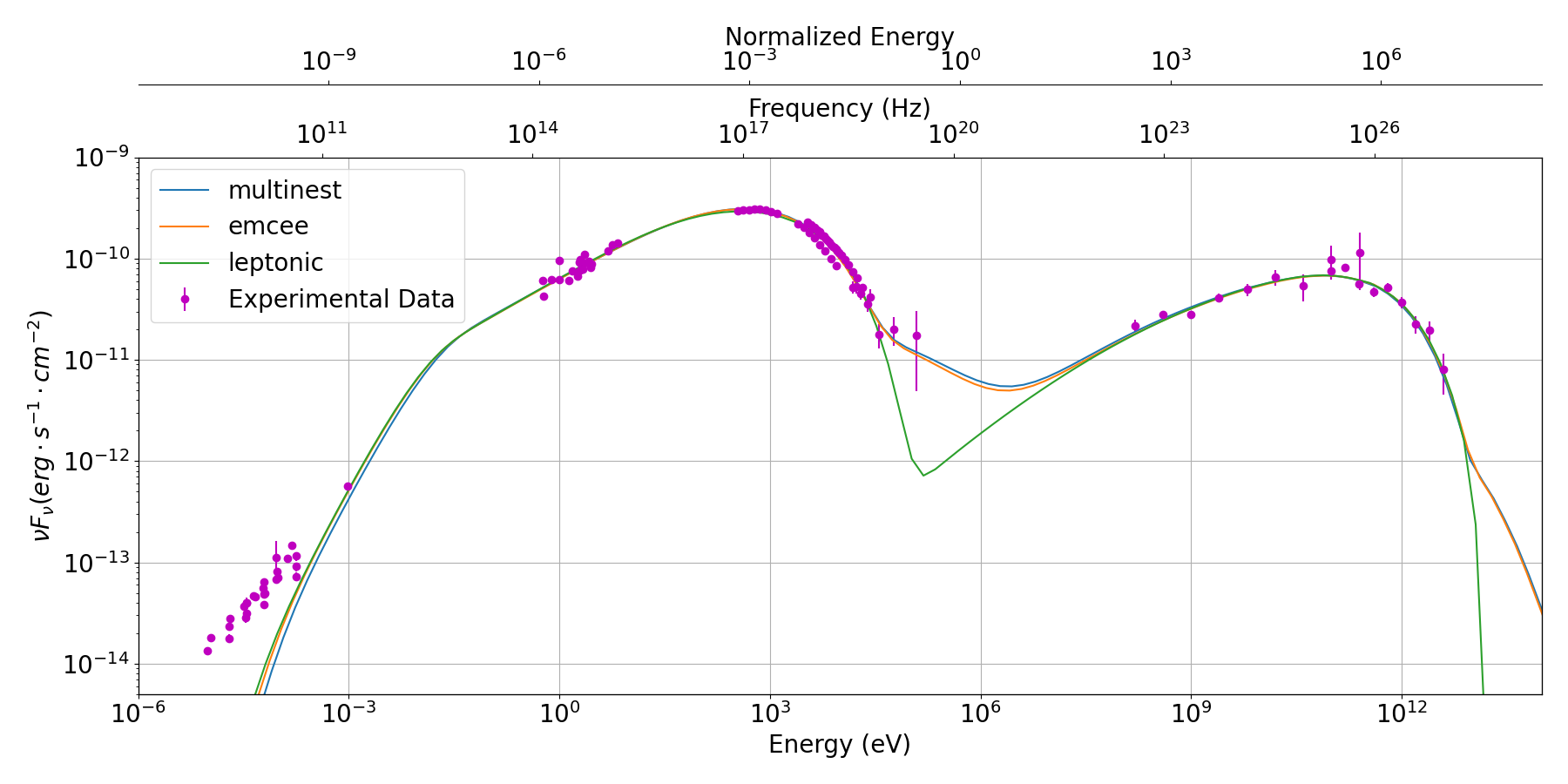}
\par\end{centering}
\caption{\label{fig:Combined-SED}Combined SED for the two statistical packages.
The blue line corresponds to the best SED obtained by \texttt{multinest},
the orange to the one obtained by \texttt{emcee} and the green one
to \texttt{multinest} with a leptonic model. The respective log likelihoods
are $-39.78$, $-39.33$ and $-42.89$.}
\end{figure*}

\begin{figure}
\begin{centering}
\includegraphics[width=1\columnwidth]{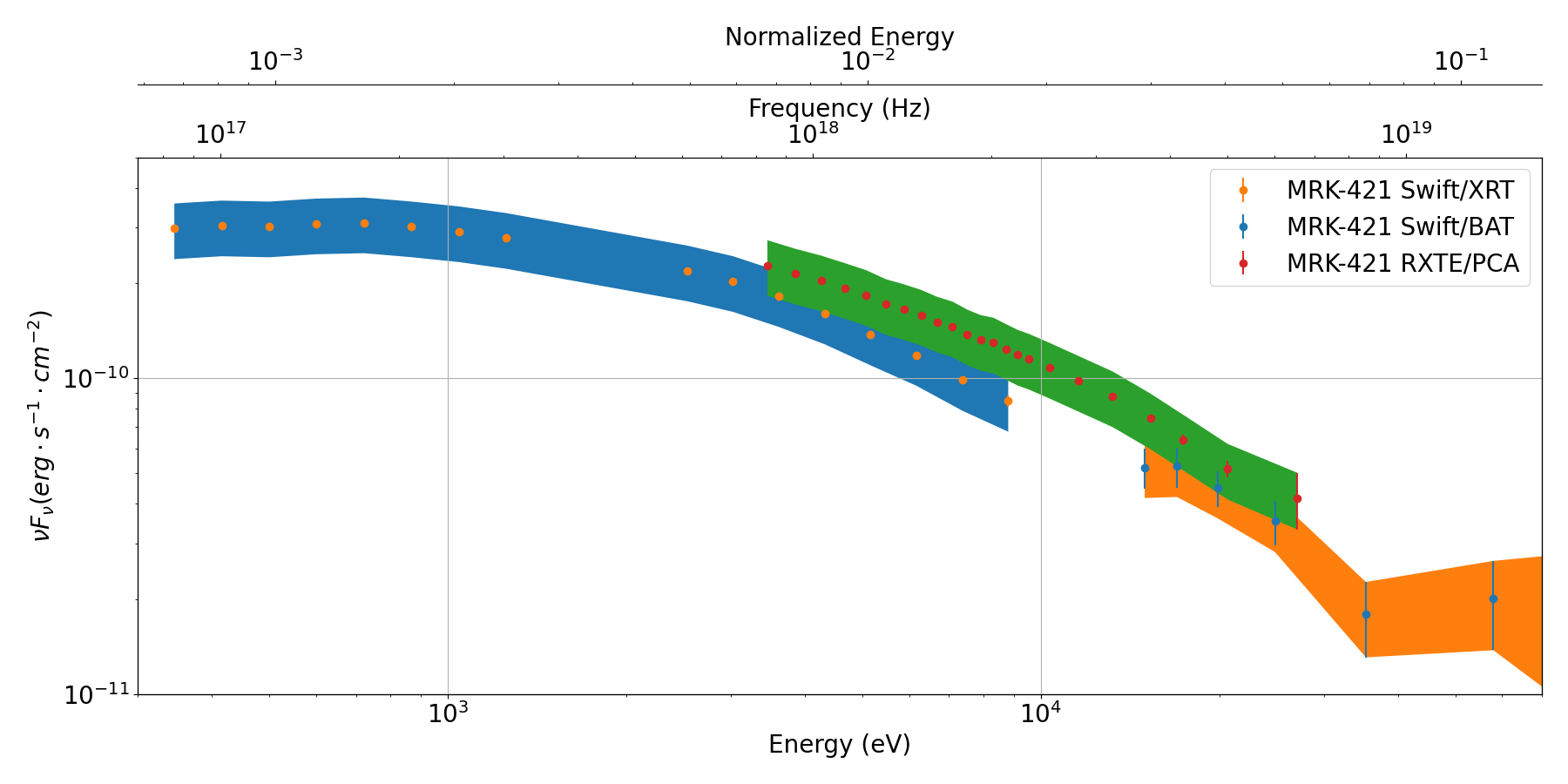}
\par\end{centering}
\caption{\label{fig:MRK-421-X-ray-data}MRK 421 X-ray data separated by instruments.
The original errors of each observation are displayed in the same
colour as the data with an errorbar. The error bands correspond to
that of taking $a=0.2$ with \eqref{error_formula}.}
\end{figure}

The full dataset that we study was gathered during a year long multi-wavelength
campaing during 2009, whose details are reported on \citet{abdo_fermi_2011}
and we reproduce it in \figref{Combined-SED}. It is important to
notice that not all the experimental points follow the same trend,
even when taking into account their errors. As an example of this
problem, we reproduce the X-rays part of the SED in \figref{MRK-421-X-ray-data},
where it can be clearly seen that there are two different trends with
the different instruments, Swift and RXTE, where the first shows a
lower flux than the second.

Similarly, observations at optical frequencies display a spectral
variability that goes beyond their (very small) observational errors.
To account for these systematic deviations from the base line physics
captured by our model, we take a conservative approach to the sizes
of the error bars of the observations (for an alternative approach
when fitting complex models to broadband transient data, see \citet{aksulu_new_2020}).
By increasing the error bars according to the prescription below,
we acknowledge that our model fit is to the generic state of the blazar
SED during the period of observation rather than to its instantaneous
value at a given time of observation for a specific instrument. We
use 
\begin{equation}
\sigma_{i}=\max\left(a\times y_{i},\sigma_{i,0}\right),\label{eq:error_formula}
\end{equation}
where $\sigma_{i}$ is the final error that we take, $a$ is a constant,
$y_{i}$ is the experimental flux and $\sigma_{i,0}$ is the experimental
error. For this analysis, we have chosen to separate the value of
$a$ by band and have set $a_{optical}=a_{\gamma-rays}=0.1$ and $a_{x-rays}=0.2$.
In \figref{MRK-421-X-ray-data} we show with bands the resulting error
bands that we obtain and see that a common trend can be more easily
fitted between the observations.

For the case of X-rays, the existence of two trends in the data had
an important impact upon the fitting procedure and its results. The
first consequence is that it creates two maxima in the likelihood
function where each one is linked to each of the trends. The existence
of more than one maximum can cause problems to fitting algorithms.
For \texttt{emcee}, as we have discussed in \subsecref{MCMC_emcee},
because it needs to have walkers in both maxima as it would be complicated
for the walkers to go from one maxima to the other, and because in
the degenerate case of having all walkers in one of the maxima the
fitting procedure would need a very high time to 'discover' the other
maxima. For \texttt{multinest} the multiples maxima are not such a
serious problem as it has the capability of discovering, separating
and analysing each of the maxima separately, although in some cases,
depending on the shape of the likelihood function this is not easy,
or even doable. In both cases, what we would find is that the running
time and the acceptance ratio of the fitting algorithms would be very
high and low respectively.

\bigskip{}

The flux detected by SMA and most of the radio observatories should
be handled carefully as they can not resolve the jet and instead detect
the integrated flux from the galaxy. To take this into account, we
have chosen to use the SMA point as an upper limit in our fits. Upper
limits can be treated in different ways in a model fit. In this case,
we have chosen to discard all model-generated SEDs that led to flux
levels above the SMA value. Because the SMA upper limit in practice
then automatically forces the SED to go below the radio upper limits
at longer wavelengths, we can safely omit these from our fitting process
without changing the outcome (confirmation of this is provided in
\subsecref{Results}).

\subsection{Configuration}{\label{subsec:Configuration}}

As we have discussed in \secref{Computational_tricks} and can be
seen in \secref{Configuration-File}, \texttt{Katu} supports a high
number of configuration options. To focus on one particular set of
models, we choose to set the plasma volume to a sphere and to allow
for all the particles to evolve. Additionally, all the initial population
distributions have ben set to simple power laws.

The background density $\left(\rho\right)$ and its ratio of protons
to electrons has been set to $10^{-26}g\cdot cm^{-3}$ and $1$ respectively.
Note that these values have no reflection on the final steady state,
as the initial distribution will be overwhelmed by the injected one.

For the electrons, we have chosen a lower cut-off for the energy range
of $2$ and higher cut-off of $\max\left(\gamma_{e,max},\frac{m_{p}}{m_{e}}\gamma_{p,max}\right)$,
where $\gamma_{e,max}$ and $\gamma_{p,max}$ are the values of the
upper cut-offs of the electron and proton injection distributions
respectively. These limits allow that electrons can cool down to very
low energies and that all the electrons that are produced through
photo-meson and the Bethe-Heitler processes can be injected in the
distribution.

Photons are also produced by many processes, so their energy limits
have to be chosen to reflect this. We have taken a lower cut-off of
$10^{-12}$ and a higher cutoff of $\gamma_{e,0,max}$, which is the
value of the higher energy cut-off for the electrons. We have also
chosen to not inject photons from the exterior of the volume, so we
set the value of the photon luminosity $\left(L_{\gamma}\right)$
as $0$.

For protons, we have taken that both, the initial and the injection
distribution have the same shape, with a lower cut-off value of $\gamma_{p,min}=10$,
a higher cut-off value of $\gamma_{p,max}=10^{6}$ and a power law
index of $2$.

We assume that the charged particles scape slower than neutral particles,
so we have set the charged to free escape ratio $\left(\sigma\right)$
as $10$.

Even though we do not know of any process that would preferentially
accelerate one kind of particle over the other, we have chosen to
set the ratio of protons to electrons $\left(\eta\right)$ as a free
variable to further explore the hadronic capabilities of our model.
In \subsecref{Leptonic-Study} we fix this ratio to $10^{-10}$ in
order to have virtually no protons and thus, be able to simulate a
fully leptonic model.

\bigskip{}

Numerically, we have chosen to set the tolerance to $10^{-6}$ and
have set the number of bins for the primary particles as: $n_{e}=n_{p}=160$
and $n_{\gamma}=128$. Experience shows that these settings provide
a good balance between numerical accuracy and reasonable convergence
times for the simulations.\bigskip{}

Finally, we are left with $8$ free variables, which are gathered
in \tabref{Parameters}, along with the respective search ranges.
$\gamma_{e,inj,min}$, $\gamma_{e,inj,max}$ and $p$ correspond to
the parameters of the injected population of electrons. $B$ $\left(\text{G}\right)$
and $\delta$ are the magnetic field and the Doppler boosting parameter
of the volume, respectively; $R$ $\left(\text{cm}\right)$ is the
radius of the sphere we simulate; and $L$ $\left(\text{erg}\cdot\text{s}^{-1}\right)$
and $\eta$ correspond to the total luminosity of the injected particles
(in the plasma frame) and its ratio of protons to electrons. As cosmological
parameters, we have chosen those of \citet{planck_collaboration_planck_2019}:
$H_{0}=67.66$, $\Omega_{m}=0.3111$ and $\Omega_{\Lambda}=0.6889$.

\begin{table}
\begin{centering}
\begin{tabular}{cc}
\toprule 
Parameter & Search Range\tabularnewline
\midrule
\midrule 
$\gamma_{e,inj,min}$ & $10-10^{4}$\tabularnewline
\midrule 
$\gamma_{e,inj,max}$ & $10^{4}-10^{10}$\tabularnewline
\midrule 
$p$ & $1-3$\tabularnewline
\midrule 
$B$ & $10^{-2}-10^{-0.5}$\tabularnewline
\midrule 
$\delta$ & $10-100$\tabularnewline
\midrule 
\multirow{1}{*}{$R$} & $10^{14}-10^{18}$\tabularnewline
\midrule 
$L$ & $10^{38}-10^{45}$\tabularnewline
\midrule 
$\eta$ & $10^{-0.1}-10^{5}$\tabularnewline
\bottomrule
\end{tabular}
\par\end{centering}
\caption{\label{tab:Parameters}Free parameters and their search ranges.}

\textcolor{red}{}
\end{table}

\subsection{Results}{\label{subsec:Results}}

The results of the two statistical packages that we have tested are
presented in \tabref{Parameter-comparisons} and \figref{Combined-SED}.
The corresponding corner plots for both packages can be seen in \figref[s]{Corner-Plot-multinest}and
\ref{fig:Corner-Plot-emcee} respectively.

\begin{table}
\begin{centering}
\begin{tabular}{ccccc}
\toprule 
\multirow{2}{*}{Parameter} & \multicolumn{2}{c}{\texttt{multinest}} & \multicolumn{2}{c}{\texttt{emcee}}\tabularnewline
 & Best & Median & Best & Median\tabularnewline
\midrule
\midrule 
$\log_{10}\left(\gamma_{e,inj,min}\right)$ & $2.50$ & $2.64_{-0.11}^{+0.16}$ & $2.55$ & $2.66_{-0.13}^{+0.17}$\tabularnewline
\midrule 
$\log_{10}\left(\gamma_{e,inj,max}\right)$ & $5.50$ & $5.61_{-0.11}^{+0.12}$ & $5.58$ & $5.61_{-0.12}^{+0.14}$\tabularnewline
\midrule 
$p$ & $2.23$ & $2.20_{-0.07}^{+0.06}$ & $2.21$ & $2.20_{-0.07}^{+0.06}$\tabularnewline
\midrule 
$\log_{10}\left(B\right)$ & $-0.96$ & $-1.03_{-0.10}^{+0.09}$ & $-1.04$ & $-1.03_{-0.10}^{+0.10}$\tabularnewline
\midrule 
$\log_{10}\left(\delta\right)$ & $1.87$ & $1.82_{-0.07}^{+0.07}$ & $1.84$ & $1.82_{-0.08}^{+0.07}$\tabularnewline
\midrule 
\multirow{1}{*}{$\log_{10}\left(R\right)$} & $15.26$ & $15.42_{-0.20}^{+0.19}$ & $15.40$ & $15.41_{-0.22}^{+0.21}$\tabularnewline
\midrule 
$\log_{10}\left(L\right)$ & $45.47$ & $45.46_{-0.92}^{+0.29}$ & $45.62$ & $45.42_{-0.97}^{+0.31}$\tabularnewline
\midrule 
$\log_{10}\left(\eta\right)$ & $3.92$ & $4.01_{-0.98}^{+0.28}$ & $4.04$ & $3.99_{-1.11}^{+0.31}$\tabularnewline
\bottomrule
\end{tabular}
\par\end{centering}
\caption{\label{tab:Parameter-comparisons}Parameter comparisons with the two
tested statistical packages.}
\end{table}

It can be clearly seen from \tabref{Parameter-comparisons} that both
approaches give very similar results for the best parameters, the
median and the $16$ and $84$ percentiles, which show the consistency
of the analysis of both packages. Likewise, from both corner plots,
\figref[s]{Corner-Plot-multinest}and \ref{fig:Corner-Plot-emcee},
we see that the posterior shape and the 2D correlations are consistent.
Furthermore, we see in \figref{Combined-SED} that both SEDs are very
close together, and indeed, their log likelihoods are $-39.78$ and
$-39.33$ for \texttt{multinest} and \texttt{emcee}, respectively.
For \texttt{multinest}, we can also quote the value of the log evidence,
which is not available for \texttt{emcee}, $\ln\left(\mathcal{Z}\right)=-65.750\pm0.232$.

The choice that we made in \subsecref{Dataset}, namely to choose
the SMA datapoint as an upper limit and ignore the rest of the radio
detections on the basis that the SMA was the most restricting one,
is correct, as none of the best fits of the SED go above any of the
radio datapoints.

In computational terms, it is complicated to compare both algorithms
due to how they handle points with null likelihood, as \texttt{multinest}
silently discards them while \texttt{emcee} treats them as any other
point; and how the first steps of \texttt{emcee} had been handled.
\texttt{multinest} has taken about a day of computations and at least
on the order of $120\,000$ evaluations of the log likelihood function,
whereas \texttt{emcee} has taken more than $400\,000$ evaluations
and more than double the time, although part of these evaluations
were done in initial short runs and due to leaving \texttt{emcee}
to work overnight without a clear stopping condition.\footnote{All the computations were done on a server with 32 available cores
with \texttt{multinest} and \texttt{emcee} calculating 8 log likelihoods
in parallel and \texttt{Katu} with 4 threads.} Even though \texttt{emcee} seems worse from the point of view of
finishing the runs, it tends to give a result for the best log likelihood
much sooner, and usually better, than \texttt{multinest}.

From \figref{Combined-SED}, it can be seen that in the range of $\sim10^{4}-10^{7}$
eV, there is a small hump between the synchrotron and inverse Compton
peaks. In our simulations, this is caused by the pairs created by
the Bethe-Heitler process \citet{mastichiadis_spectral_2005,petropoulou_betheheitler_2015},
highlighting the need for protons in order to completely fit the SED.

\citet{abdo_fermi_2011} estimate the luminosity of the blazar between
$2.6\times10^{46}$ and $1.2\times10^{47}$. Taking that the luminosity
from the jet frame to the blazar frame transforms as $L_{blazar}\approx\delta^{2}L_{jet}/2$,
we obtain $L_{blazar,multinest}\approx6.44\times10^{48}$ and $L_{blazar,emcee}\approx9.98\times10^{48}$.
Both values are clearly above the estimation of \citet{abdo_fermi_2011},
an issue shared with other lepto hadronic models \citet{liodakis_proton_2020}.

\subsection{Model selection using Bayesian inference: leptonic versus lepto-hadronic
models}{\label{subsec:Leptonic-Study}}

To verify that the small hump that appears between the peaks in \figref{Combined-SED}
is indeed caused by the presence of protons and to compare our results
against a purely leptonic model, we choose to repeat the statistical
analysis of Mrk 421's data with \texttt{multinest} but setting $\eta=10^{-10}$.
This ratio of injected protons to electrons means that, effectively,
the hadronic processes have no impact upon the final fit. The choice
of \texttt{multinest} over \texttt{emcee} is to allow us to obtain
a value for the evidence that we can compare against the evidence
for the hadronic model.

As we can see in \figref{Combined-SED}, the best SED that the leptonic
model can produce is unable to fit the high energy part of the X-ray
observations, and this is indeed reflected in the best log likelihood
that it can achieve: $-42.89$. We also find that the log evidence
has a value of $\ln\left(\mathcal{Z}\right)=-67.685\pm0.231$.

Assuming that both models have the same a priori probability, we find
that the hadronic model is favoured as:
\[
\ln\left(R\right)\approx1.935\pm0.327.
\]

The corresponding Bayes' Factor is $6.92_{-1.93}^{+2.68}$. This value
is consistent with the qualification \emph{substantial} for the evidence
in favour of the lepto-hadronic model, as discussed in \subsecref{Nomenclature}.
Note that this measure already accounts for the different number of
parameters between the two models.

Lepto-hadronic models can successfully explain some of the features
of blazar SEDs, such as the gamma ray peak being caused by neutral
pion decay or proton synchrotron \citet{mannheim_proton_1993,cerruti_hadronic_2015}.
But, leptonic models can also fit this peak with inverse compton scattering
of synchrotron photons, and the energy requirements for the proton
population are usually above the Eddington luminosity \citet{liodakis_proton_2020}.
Some characteristics of the SEDs can only be explained by a hadronic
population, such as a population of pairs produced by the Bethe-Heitler
pair production process, or the presence of neutrinos \citet{mastichiadis_spectral_2005,petropoulou_betheheitler_2015,petropoulou_photohadronic_2015,rodrigues_blazars_2019,rodrigues_multi-wavelength_2020}.
However, to our knowledge, this is the first time that the dichotomy
of leptonic vs. lepto-hadronic modelling has been explored from a
fully bayesian point of view using kinetic equation modelling.

\section{Summary}{\label{sec:Conclusions}}

In this work we introduce the code \texttt{Katu}, for simulating the
evolution of plasmas according to a set of kinetic equation for a
series of particles. We have tried to capture as much of the physics
as possible by adding diverse processes in a detailed way. We have
tried to make it as performant as possible, on the mathematical front
by adapting the kinetic equation to the species of particle and separating
neutral from charged particles, and on the computational front by
takind advantage of multi-threading and by calculating as much as
possible up-front.

Given that \texttt{Katu} is fast, we are able to tie it to two commonly
used statistical packages that can perform bayesian analysis of models,
\texttt{emcee} and \texttt{multinest}. \texttt{emcee} is a python
package that performs bayesian analysis by using affine invariant
samplers for MCMC and \texttt{multinest} is a \texttt{Fortran} package,
which we employ using \texttt{pymultinest}, which implements Elliptical
Nested Sampling.

As an example of its usage, we show how \texttt{Katu} and both wrappers
can be used to analyse the well known blazar Mrk 421 obtaining, as
expected, that both statistical packages report similar values for
the best, median and $16$ and $84$ percentiles of the posterior
of the parameters.

Additionally, we perform the test of exploring a purely leptonic model,
finding that there is substantial evidence to reject it in favour
of a lepto-hadronic model.

The code \texttt{Katu} was also employed for the study of TXS 0506+056,
as described in \citet{jimenez-fernandez_bayesian_2020}.

\section*{Acknowledgements}

The authors would like to thank Matteo Cerruti and Anatoli Fedynitch
for the valuable discussions about their models and Petar Mimica for
hosting them during their meetings at the University of Valencia.
B.J.F. would also like to thank the late Cosmo the Cat, Jinx, Loki
and Mina the Cats and Lottie the Dog for their support. B.J.F. acknowledges
funding from STFC through the joint STFC-IAC scholarship program.
H.J.v.E. acknowledges partial support by the European Union Horizon
2020 Programme under the AHEAD2020 project (grant agreement number
871158).

\bibliographystyle{mnras}
\bibliography{bibliography/biblio}

\begin{thebibliography}{}
\makeatletter
\relax
\def\mn@urlcharsother{\let\do\@makeother \do\$\do\&\do\#\do\^\do\_\do\%\do\~}
\def\mn@doi{\begingroup\mn@urlcharsother \@ifnextchar [ {\mn@doi@}
  {\mn@doi@[]}}
\def\mn@doi@[#1]#2{\def\@tempa{#1}\ifx\@tempa\@empty \href
  {http://dx.doi.org/#2} {doi:#2}\else \href {http://dx.doi.org/#2} {#1}\fi
  \endgroup}
\def\mn@eprint#1#2{\mn@eprint@#1:#2::\@nil}
\def\mn@eprint@arXiv#1{\href {http://arxiv.org/abs/#1} {{\tt arXiv:#1}}}
\def\mn@eprint@dblp#1{\href {http://dblp.uni-trier.de/rec/bibtex/#1.xml}
  {dblp:#1}}
\def\mn@eprint@#1:#2:#3:#4\@nil{\def\@tempa {#1}\def\@tempb {#2}\def\@tempc
  {#3}\ifx \@tempc \@empty \let \@tempc \@tempb \let \@tempb \@tempa \fi \ifx
  \@tempb \@empty \def\@tempb {arXiv}\fi \@ifundefined
  {mn@eprint@\@tempb}{\@tempb:\@tempc}{\expandafter \expandafter \csname
  mn@eprint@\@tempb\endcsname \expandafter{\@tempc}}}

\bibitem[\protect\citeauthoryear{Abdo et~al.,}{Abdo
  et~al.}{2011}]{abdo_fermi_2011}
Abdo A.~A.,  et~al., 2011, \mn@doi [The Astrophysical Journal]
  {10.1088/0004-637X/736/2/131}, 736, 131

\bibitem[\protect\citeauthoryear{Aksulu, Wijers, van Eerten  \&
  van der Horst}{Aksulu et~al.}{2020}]{aksulu_new_2020}
Aksulu M.~D.,  Wijers R. A. M.~J.,  van Eerten H.~J.,   van der Horst A.~J.,
   2020, \mn@doi [Monthly Notices of the Royal Astronomical Society]
  {10.1093/mnras/staa2297}, 497, 4672

\bibitem[\protect\citeauthoryear{Blumenthal}{Blumenthal}{1970}]{blumenthal_energy_1970}
Blumenthal G.~R.,  1970, \mn@doi [Physical Review D] {10.1103/PhysRevD.1.1596},
  1, 1596

\bibitem[\protect\citeauthoryear{B\"ottcher \& Schlickeiser}{B\"ottcher \&
  Schlickeiser}{1997}]{bottcher_pair_1997}
B\"ottcher M.,  Schlickeiser R.,  1997, Astronomy \& Astrophysics, p.~5

\bibitem[\protect\citeauthoryear{{B\"ottcher}, {Harris}  \&
  {Krawczynski}}{{B\"ottcher} et~al.}{2012}]{Bottcher_Book}
{B\"ottcher} M.,  {Harris} D.~E.,   {Krawczynski} H.,  2012, {Relativistic Jets
  from Active Galactic Nuclei}

\bibitem[\protect\citeauthoryear{Cerruti, Zech, Boisson  \& Inoue}{Cerruti
  et~al.}{2015}]{cerruti_hadronic_2015}
Cerruti M.,  Zech A.,  Boisson C.,   Inoue S.,  2015, \mn@doi [Monthly Notices
  of the Royal Astronomical Society] {10.1093/mnras/stu2691}, 448, 910

\bibitem[\protect\citeauthoryear{Collaboration et~al.,}{Collaboration
  et~al.}{2019}]{planck_collaboration_planck_2019}
Collaboration P.,  et~al., 2019, arXiv:1807.06209 [astro-ph]

\bibitem[\protect\citeauthoryear{Crusius \& Schlickeiser}{Crusius \&
  Schlickeiser}{1986}]{crusius_synchrotron_1986}
Crusius A.,  Schlickeiser R.,  1986, Astronomy \& Astrophysics, 164, L16

\bibitem[\protect\citeauthoryear{Dimitrakoudis, Mastichiadis, Protheroe  \&
  Reimer}{Dimitrakoudis et~al.}{2012}]{dimitrakoudis_time-dependent_2012}
Dimitrakoudis S.,  Mastichiadis A.,  Protheroe R.~J.,   Reimer A.,  2012,
  \mn@doi [Astronomy \& Astrophysics] {10.1051/0004-6361/201219770}, 546, A120

\bibitem[\protect\citeauthoryear{Feroz \& Hobson}{Feroz \&
  Hobson}{2008}]{feroz_multimodal_2008}
Feroz F.,  Hobson M.~P.,  2008, \mn@doi [Monthly Notices of the Royal
  Astronomical Society] {10.1111/j.1365-2966.2007.12353.x}, 384, 449

\bibitem[\protect\citeauthoryear{Feroz, Hobson  \& Bridges}{Feroz
  et~al.}{2009}]{feroz_multinest:_2009}
Feroz F.,  Hobson M.~P.,   Bridges M.,  2009, \mn@doi [Monthly Notices of the
  Royal Astronomical Society] {10.1111/j.1365-2966.2009.14548.x}, 398, 1601

\bibitem[\protect\citeauthoryear{Feroz, Hobson, Cameron  \& Pettitt}{Feroz
  et~al.}{2013}]{feroz_importance_2013}
Feroz F.,  Hobson M.~P.,  Cameron E.,   Pettitt A.~N.,  2013, arXiv:1306.2144
  [astro-ph, physics:physics, stat]

\bibitem[\protect\citeauthoryear{Foreman-Mackey, Hogg, Lang  \&
  Goodman}{Foreman-Mackey et~al.}{2013}]{foreman-mackey_emcee:_2013}
Foreman-Mackey D.,  Hogg D.~W.,  Lang D.,   Goodman J.,  2013, \mn@doi
  [Publications of the Astronomical Society of the Pacific] {10.1086/670067},
  125, 306

\bibitem[\protect\citeauthoryear{Gao, Pohl  \& Winter}{Gao
  et~al.}{2017}]{gao_direct_2017}
Gao S.,  Pohl M.,   Winter W.,  2017, \mn@doi [The Astrophysical Journal]
  {10.3847/1538-4357/aa7754}, 843, 109

\bibitem[\protect\citeauthoryear{Gao, Fedynitch, Winter  \& Pohl}{Gao
  et~al.}{2019}]{gao_modelling_2019}
Gao S.,  Fedynitch A.,  Winter W.,   Pohl M.,  2019, \mn@doi [Nature Astronomy]
  {10.1038/s41550-018-0610-1}, 3, 88

\bibitem[\protect\citeauthoryear{Ghisellini}{Ghisellini}{2013}]{ghisellini_radiative_2013}
Ghisellini G.,  2013, \mn@doi [arXiv:1202.5949 [astro-ph]]
  {10.1007/978-3-319-00612-3}, 873

\bibitem[\protect\citeauthoryear{Goodman \& Weare}{Goodman \&
  Weare}{2010}]{goodman_ensemble_2010}
Goodman J.,  Weare J.,  2010, \mn@doi [Communications in Applied Mathematics
  and Computational Science] {10.2140/camcos.2010.5.65}, 5, 65

\bibitem[\protect\citeauthoryear{H\"ummer, R\"uger, Spanier  \&
  Winter}{H\"ummer et~al.}{2010}]{hummer_simplified_2010}
H\"ummer S.,  R\"uger M.,  Spanier F.,   Winter W.,  2010, \mn@doi [The
  Astrophysical Journal] {10.1088/0004-637X/721/1/630}, 721, 630

\bibitem[\protect\citeauthoryear{Jim\'enez-Fern\'andez \& van\
  Eerten}{Jim\'enez-Fern\'andez \& van\
  Eerten}{2020}]{jimenez-fernandez_bayesian_2020}
Jim\'enez-Fern\'andez B.,  van\ Eerten H.~J.,  2020, \mn@doi [Monthly Notices
  of the Royal Astronomical Society] {10.1093/mnras/staa3163}, 500, 3613

\bibitem[\protect\citeauthoryear{Jones}{Jones}{1968}]{jones_calculated_1968}
Jones F.~C.,  1968, \mn@doi [Physical Review] {10.1103/PhysRev.167.1159}, 167,
  1159

\bibitem[\protect\citeauthoryear{Kass \& Raftery}{Kass \&
  Raftery}{1995}]{bayes_factors}
Kass R.~E.,  Raftery A.~E.,  1995, Journal of the American Statistical
  Association, 90, 773

\bibitem[\protect\citeauthoryear{Keivani et~al.,}{Keivani
  et~al.}{2018}]{keivani_multimessenger_2018}
Keivani A.,  et~al., 2018, \mn@doi [The Astrophysical Journal]
  {10.3847/1538-4357/aad59a}, 864, 84

\bibitem[\protect\citeauthoryear{Kelner \& Aharonian}{Kelner \&
  Aharonian}{2008}]{kelner_energy_2008}
Kelner S.~R.,  Aharonian F.~A.,  2008, \mn@doi [Physical Review D]
  {10.1103/PhysRevD.78.034013}, 78

\bibitem[\protect\citeauthoryear{Lightman \& Zdziarski}{Lightman \&
  Zdziarski}{1987}]{lightman_pair_1987}
Lightman A.~P.,  Zdziarski A.~A.,  1987, \mn@doi [The Astrophysical Journal]
  {10.1086/165485}, 319, 643

\bibitem[\protect\citeauthoryear{Liodakis \& Petropoulou}{Liodakis \&
  Petropoulou}{2020}]{liodakis_proton_2020}
Liodakis I.,  Petropoulou M.,  2020, \mn@doi [The Astrophysical Journal]
  {10.3847/2041-8213/ab830a}, 893, L20

\bibitem[\protect\citeauthoryear{Lipari, Lusignoli  \& Meloni}{Lipari
  et~al.}{2007}]{lipari_flavor_2007}
Lipari P.,  Lusignoli M.,   Meloni D.,  2007, \mn@doi [Physical Review D]
  {10.1103/PhysRevD.75.123005}, 75

\bibitem[\protect\citeauthoryear{Mannheim}{Mannheim}{1993}]{mannheim_proton_1993}
Mannheim K.,  1993, arXiv:astro-ph/9302006

\bibitem[\protect\citeauthoryear{Mastichiadis \& Kirk}{Mastichiadis \&
  Kirk}{1992}]{mastichiadis_x-ray_1992}
Mastichiadis A.,  Kirk J.~G.,  1992, Nature, 360, 135

\bibitem[\protect\citeauthoryear{Mastichiadis \& Kirk}{Mastichiadis \&
  Kirk}{1995}]{mastichiadis_self-consistent_1995}
Mastichiadis A.,  Kirk J.~G.,  1995, Astronomy \& Astrophysics, 295, 613

\bibitem[\protect\citeauthoryear{Mastichiadis, Protheroe  \& Kirk}{Mastichiadis
  et~al.}{2005}]{mastichiadis_spectral_2005}
Mastichiadis A.,  Protheroe R.~J.,   Kirk J.~G.,  2005, \mn@doi [Astronomy \&
  Astrophysics] {10.1051/0004-6361:20042161}, 433, 765

\bibitem[\protect\citeauthoryear{Petropoulou \& Mastichiadis}{Petropoulou \&
  Mastichiadis}{2015}]{petropoulou_betheheitler_2015}
Petropoulou M.,  Mastichiadis A.,  2015, \mn@doi [Monthly Notices of the Royal
  Astronomical Society] {10.1093/mnras/stu2364}, 447, 36

\bibitem[\protect\citeauthoryear{Petropoulou, Dimitrakoudis, Padovani,
  Mastichiadis  \& Resconi}{Petropoulou
  et~al.}{2015}]{petropoulou_photohadronic_2015}
Petropoulou M.,  Dimitrakoudis S.,  Padovani P.,  Mastichiadis A.,   Resconi
  E.,  2015, \mn@doi [Monthly Notices of the Royal Astronomical Society]
  {10.1093/mnras/stv179}, 448, 2412

\bibitem[\protect\citeauthoryear{Petropoulou, Oikonomou, Mastichiadis, Murase,
  Padovani, Vasilopoulos  \& Giommi}{Petropoulou
  et~al.}{2020}]{petropoulou_comprehensive_2020}
Petropoulou M.,  Oikonomou F.,  Mastichiadis A.,  Murase K.,  Padovani P.,
  Vasilopoulos G.,   Giommi P.,  2020, \mn@doi [The Astrophysical Journal]
  {10.3847/1538-4357/aba8a0}, 899, 113

\bibitem[\protect\citeauthoryear{Rodrigues}{Rodrigues}{2019}]{rodrigues_blazars_2019}
Rodrigues X.,  2019, PhD thesis, Humboldt University, \url
  {https://ui.adsabs.harvard.edu/abs/2019PhDT........87R/abstract}

\bibitem[\protect\citeauthoryear{Rodrigues, Garrappa, Gao, Paliya, Franckowiak
  \& Winter}{Rodrigues et~al.}{2020}]{rodrigues_multi-wavelength_2020}
Rodrigues X.,  Garrappa S.,  Gao S.,  Paliya V.~S.,  Franckowiak A.,   Winter
  W.,  2020, arXiv:2009.04026 [astro-ph]

\bibitem[\protect\citeauthoryear{Rybicki \& Lightman}{Rybicki \&
  Lightman}{1979}]{rybicki_radiative_1979}
Rybicki G.~B.,  Lightman A.~P.,  1979, Radiative processes in astrophysics.
Wiley, New York

\bibitem[\protect\citeauthoryear{Svensson}{Svensson}{1982}]{svensson_pair_1982}
Svensson R.,  1982, \mn@doi [The Astrophysical Journal] {10.1086/160081}, 258,
  321

\makeatother
\end{thebibliography}

\appendix
\onecolumn

\section{Detailed Description of the processes}{\label{sec:Detailed-Description}}

In this appendix we try to give a detailed description of the equations
that we implement in our model, along with the references. The reader
must notice two things. First, we have adapted the notation to make
it consistent with our conventions through the text. Additionally,
some equations have been rearranged into equivalent forms that are
a bit simpler. Second, our numerical implementation, even though it
follows these equations as closely as possible, has some quirks that
the reader and anyone implementing the model should be aware of.

Note also that we do not explicitly add anything about the losses
of particles due to decay because this term is explicitly tracked
in the kinetic equation, \ref{eq:kinetic_equation}.

In the first place, some of the equations are physically valid in
limits such as $\gamma\rightarrow1$, $\varepsilon\rightarrow0$,
$1-\beta\rightarrow0$, etc. but numerically they either become unstable
or they fall into regimes where 'catastrophic loss of precision' could
happen. In these cases, the special cases have been pulled out and
are treated separately or are accounted for by using approximations.

In the second case, in general, the integrals are taken in logarithmic
space in energy, as that makes the bins equally sized and in general
the energy term that is introduced can be simplified, which also simplifies
the equations. Numerically, they are performed using the trapezium
rule, which gives a good precision taking into account the errors
already introduced by the approximations in the models and the final
data we compare with. Likewise, the derivatives are calculated using
a first order scheme, although in log-log space, as they are much
more stable this way. 

\subsection{Synchrotron Emission and Absorption}

In our model, all the charged particles create synchrotron emission,
have synchrotron losses and cause synchrotron absorption. For this,
we model the emission as:
\begin{equation}
j_{\nu}=\frac{1}{4\pi}\int_{1}^{\infty}n\left(\gamma\right)P_{\nu}\left(\gamma\right)d\gamma,
\end{equation}
where $P_{\nu}\left(\gamma\right)$ is the energy emitted by a particle
with energy $\gamma$ at frequency $\nu$ and has the form \citep{crusius_synchrotron_1986}:
\begin{equation}
P_{\nu}\left(\gamma\right)=\frac{\sqrt{3}}{4\pi}\frac{q^{3}B}{mc^{2}}xCS\left(x\right),
\end{equation}
where $q$ and $m$ are the charge and the mass of the emitting particle,
$B$ is the magnetic field strength, $x$ is the frequency normalized
to a critical frequency $x=\nu/\nu_{c}$ with
\begin{equation}
\nu_{c}\coloneqq\nu_{0}\gamma^{2}\coloneqq\frac{3}{4\pi}\frac{qB}{mc}\gamma^{2},
\end{equation}
and $CS\left(x\right)$ is a function given in terms of Whittaker
functions:
\begin{equation}
CS\left(x\right)=W_{0,\frac{4}{3}}\left(x\right)W_{0,\frac{1}{3}}\left(x\right)-W_{\frac{1}{2},\frac{5}{6}}\left(x\right)W_{-\frac{1}{2},\frac{5}{6}}\left(x\right).
\end{equation}

The loss of energy of the emitting particles is modelled as:
\begin{equation}
\frac{\partial\gamma}{\partial t}=-\frac{4}{3}c\sigma_{T}\frac{u_{B}}{m_{e}c^{2}}\left(\frac{m_{e}}{m}\right)^{3}\gamma^{2},
\end{equation}
where $\sigma_{T}$ is the Thomson cross-section and $u_{B}$ is the
energy density of the magnetic field.

The absorption of photons is modelled as:
\begin{equation}
\alpha_{\nu}=-\frac{1}{8\pi m\nu^{2}}\int_{1}^{\infty}\gamma^{2}\frac{\partial}{\partial\gamma}\left[\frac{n}{\gamma^{2}}\right]P_{\nu}\left(\gamma\right)d\gamma.
\end{equation}

\subsection{Inverse Compton}

Photons interact with energetic electrons due to the inverse Compton
process, which we model as:
\begin{equation}
j_{\nu}\left(\varepsilon_{s}\right)=\frac{\hbar\varepsilon_{s}}{2}\int_{1}^{\infty}n_{e}\left(\gamma_{e}\right)\int_{0}^{\infty}n_{\gamma}\left(\varepsilon\right)g_{u,d}\left(\gamma_{e},\varepsilon,\varepsilon_{s}\right)d\varepsilon d\gamma_{e},
\end{equation}
where $\varepsilon$ is the initial energy of the photon, $\varepsilon_{s}$
is the energy after the scattering, $n_{e}$ is the electron population,
$n_{\gamma}$ is the photon population and $g_{u,d}\left(\gamma,\varepsilon,\varepsilon_{s}\right)$
are two functions referring to the upscattering and downscattering
of photons which where calculated by Jones \citep[equations 40 and 44]{jones_calculated_1968}:
\begin{align}
g_{u}\left(\gamma,\varepsilon,\varepsilon_{s}\right) & =\frac{3c\sigma_{T}}{4\gamma^{2}\varepsilon}\left[2q_{u}\ln q_{u}+\left(1+2q_{u}\right)\left(1-q_{u}\right)+\frac{\varepsilon_{s}^{2}}{\gamma\left(\gamma-\varepsilon_{s}\right)}\frac{1-q_{u}}{2}\right]\quad\text{if }q_{u}\leq1\leq\frac{\varepsilon_{s}}{\varepsilon},\,\varepsilon_{s}<\gamma,\\
g_{d}\left(\gamma,\varepsilon,\varepsilon_{s}\right) & =\frac{3c\sigma_{T}}{16\gamma^{4}\varepsilon}\left[\left(q_{d}-1\right)\left(1+\frac{2}{q_{d}}\right)-2\ln q_{d}\right]\qquad\qquad\qquad\qquad\qquad\text{if }\frac{1}{q_{d}}\leq1\leq\frac{\varepsilon}{\varepsilon_{s}},
\end{align}
with
\begin{align}
q_{u} & =\frac{\varepsilon_{s}}{4\varepsilon\gamma\left(\gamma-\varepsilon_{s}\right)}, & q_{d} & =\frac{4\gamma{{}^2}\varepsilon_{s}}{\varepsilon}.
\end{align}

The loss term for photons can be obtained as:
\begin{equation}
\frac{\partial n_{\gamma}\left(\varepsilon\right)}{\partial t}=-n_{\gamma}\left(\varepsilon\right)\int_{1}^{\infty}n_{e}\left(\gamma_{e}\right)\mathcal{R}\left(\gamma_{e},\varepsilon\right)d\gamma_{e},\label{eq:IC_losses-1}
\end{equation}
where $\mathcal{R}\left(\gamma_{e},\varepsilon\right)$ is the reaction
rate of electrons of energy $\gamma_{e}$ with photons of energy $\varepsilon$
and can be found from integrating the Klein-Nishina cross section.

\begin{align}
\mathcal{R}\left(\gamma_{e},\varepsilon\right) & =\frac{3}{16}c\sigma_{T}\frac{1}{\beta\gamma_{e}^{2}\varepsilon^{2}}\left[-\frac{x}{4}+\frac{1}{4\left(x+1\right)}+\frac{\left(x^{2}+9x+8\right)}{2x}\ln\left(1+x\right)+2\Li_{2}\left(-x\right)\right]_{x_{M}}^{x_{m}}
\end{align}
where:
\begin{align}
x_{m} & =2\varepsilon\gamma_{e}\left(1+\beta\right), & x_{M} & =2\varepsilon\gamma_{e}\left(1-\beta\right),
\end{align}
and $\Li_{2}$ the dilogarithm.

Following \citet{Bottcher_Book}, we use the simplified prescription
for the electron energy losses in inverse Compton scattering:
\begin{equation}
\frac{\partial\gamma}{\partial t}=-\frac{4}{3}c\sigma_{T}\frac{u_{\gamma}}{m_{e}c^{2}}\gamma^{2},
\end{equation}
where $u_{\gamma}$ is the energy density of the photon field:
\begin{equation}
u_{\gamma}=\int_{0}^{\infty}\varepsilon n_{\gamma}\left(\varepsilon\right)d\varepsilon.
\end{equation}

\subsection{Two photon Pair Production}

Photons can interact among themselves creating electron and positron
pairs, which we model according to \citep{bottcher_pair_1997}.

Let us define
\begin{align}
E & =\varepsilon_{1}+\varepsilon_{2}, & c_{\pm} & =\left(\varepsilon_{1,2}-\gamma\right)^{2}-1,\\
a & =\gamma\left(E-\gamma\right)+1, & d_{\pm} & =\varepsilon_{1,2}^{2}+\varepsilon_{1}\varepsilon_{2}\pm\gamma\left(\varepsilon_{2}-\varepsilon_{1}\right),
\end{align}
where $\varepsilon_{1,2}$ are the energies of the initial photons
and $\gamma$ is the energy of the final electron or positron.

Then we can obtain the rate of created particles as:
\begin{equation}
\frac{\partial n_{e^{\pm}}\left(\gamma\right)}{\partial t}=\frac{3}{4}\sigma_{T}c\int_{0}^{\infty}d\varepsilon_{1}\frac{n_{\gamma}\left(\varepsilon_{1}\right)}{\varepsilon_{1}^{2}}\int_{\max\left(\frac{1}{\varepsilon_{1}},\gamma+1-\varepsilon_{1}\right)}^{\infty}d\varepsilon_{2}\frac{n_{\gamma}\left(\varepsilon_{2}\right)}{\varepsilon_{2}^{2}}R\left(\varepsilon_{1},\varepsilon_{2},\gamma\right),
\end{equation}
where
\begin{equation}
R\left(\varepsilon_{1},\varepsilon_{2},\gamma\right)=\left[\frac{\sqrt{E^{2}-4\varepsilon'^{2}}}{4}+H_{+}+H_{-}\right]_{\varepsilon_{min}'}^{\varepsilon_{max}'}.
\end{equation}
The values for $\varepsilon'$ are given by energetic and kinematic
reasoning:
\begin{align}
\varepsilon_{max}' & =\min\left(\sqrt{\varepsilon_{1}\varepsilon_{2}},\varepsilon'^{\star}\right), & \varepsilon_{min}' & =\max\left(1,\varepsilon'^{\dagger}\right),
\end{align}
where
\begin{equation}
\left(\varepsilon'^{\star,\dagger}\right)^{2}=\frac{a\pm\sqrt{a^{2}-E^{2}}}{2}.
\end{equation}

The function $H_{\pm}$ is given depending on the value of $c_{\pm}$:

\begin{align}
H_{\pm} & =\begin{cases}
\frac{\varepsilon'}{8\sqrt{\varepsilon_{1}\varepsilon_{2}+c_{\pm}\varepsilon'^{2}}}\left[2\left(\varepsilon'^{2}+\frac{1}{\varepsilon'^{2}}\right)+2\frac{\varepsilon_{1}\varepsilon_{2}-1}{c_{\pm}}+\frac{2c_{\pm}-d_{\pm}}{\varepsilon_{1}\varepsilon_{2}}\right]+\frac{I_{\pm}}{4}\left(2-\frac{\varepsilon_{1}\varepsilon_{2}-1}{c_{\pm}}\right) & \text{if }c_{\pm}\neq0\\
\frac{1}{\left(\varepsilon_{1}\varepsilon_{2}\right)^{3/2}}\left(\frac{\varepsilon'^{3}}{12}-\frac{\varepsilon'd_{\pm}}{8}\right)+\frac{1}{\sqrt{\varepsilon_{1}\varepsilon_{2}}}\left(\frac{\varepsilon'^{3}}{6}+\frac{\varepsilon'}{2}+\frac{1}{4\varepsilon'}\right) & \text{if }c_{\pm}=0
\end{cases},
\end{align}
with
\begin{equation}
I_{\pm}=\begin{cases}
\frac{1}{\sqrt{c_{\pm}}}\log\left(\varepsilon'\sqrt{c_{\pm}}+\sqrt{\varepsilon_{1}\varepsilon_{2}+c_{\pm}\varepsilon'^{2}}\right) & \text{if }c_{\pm}>0\\
\frac{1}{\sqrt{-c_{\pm}}}\arcsin\left(\varepsilon'\sqrt{-\frac{c_{\pm}}{\varepsilon_{1}\varepsilon_{2}}}\right) & \text{if }c_{\pm}<0
\end{cases}.
\end{equation}

\bigskip{}

Photon losses are modelled according to the fully integrated cross
section in a similar way to the photon loss term for IC.

\begin{align}
\mathcal{R}\left(x\right) & =\frac{3\sigma_{T}}{4x^{2}}\left\{ a-2xa+\Li_{2}\left(\frac{1-a}{2}\right)-\Li_{2}\left(\frac{1+a}{2}\right)-\atanh\left(a\right)\left[-\ln\left(4x\right)-\frac{1}{x}-2x+2\right]\right\} ,
\end{align}
where $x=\varepsilon_{1}\varepsilon_{2}$ and $a=\sqrt{1-\frac{1}{x}}$.

\subsection{Pair Annihilation}

Just like a pair of photons can interact to produce a pair electron-positron,
the opposite process can happen where electrons and positrons annihilate
among themselves to produce a pair of photons. We model this process
following \citet{svensson_pair_1982}.

Let us start by defining the auxiliary variables:
\begin{align}
\gamma_{cm}^{*2} & \coloneqq\varepsilon\left(\gamma_{+}+\gamma_{-}-\varepsilon\right), & d_{\pm} & \coloneqq\gamma_{\mp}\left(\gamma_{+}+\gamma_{-}\right)\pm\varepsilon\left(\gamma_{+}-\gamma_{-}\right),\\
c_{\pm} & \coloneqq\left(\gamma_{\mp}-\varepsilon\right)^{2}-1, & u_{\pm} & \coloneqq\sqrt{\gamma_{cm}^{*2}+c_{\pm}\gamma_{cm}^{2}},
\end{align}
note that these definitions are similar but different to the case
of pair production. In particular, $c_{\pm}$ and $d_{\pm}$ go with
$\gamma_{\mp}$ not the other way around.

\citet{svensson_pair_1982} defines the spectral emissivity as:
\begin{equation}
\frac{\partial n_{\gamma}\left(\varepsilon\right)}{\partial t}=\int_{\gamma_{+,min}}^{\gamma_{+,max}}n_{e^{+}}\left(\gamma_{+}\right)\int_{\gamma_{-,min}}^{\gamma_{-,max}}n_{e^{-}}\left(\gamma_{-}\right)\overline{v\frac{d\sigma}{d\varepsilon}}\left(\varepsilon,\gamma_{+},\gamma_{-}\right)d\gamma_{-}d\gamma_{+}
\end{equation}
where $\overline{v\frac{d\sigma}{d\varepsilon}}$ is the angle averaged
emissivity and the limits $\gamma_{\pm,min}$ and $\gamma_{\pm,max}$
are defined using the auxiliary functions:
\begin{align}
\gamma_{A}\left(\varepsilon\right) & \coloneqq\frac{4\varepsilon^{2}+1}{4\varepsilon}, & F_{\pm}\left(\varepsilon,\gamma\right) & \coloneqq2\varepsilon-\gamma\left(1\pm\beta\right),\\
\gamma_{B}\left(\varepsilon\right) & \coloneqq\frac{2\varepsilon^{2}-2\varepsilon+1}{2\varepsilon-1}, & \gamma^{\pm}\left(\varepsilon,\gamma\right) & \coloneqq\frac{1}{2}\left(F_{\pm}+\frac{1}{F_{\pm}}\right),
\end{align}
which give the integration limits as:
\begin{align}
\gamma_{\pm,max} & =\infty,\\
\gamma_{+,min} & =\begin{cases}
1 & \text{if }\varepsilon>\frac{1}{2}\\
\gamma_{A}\left(\varepsilon\right) & \text{if }\varepsilon\leq\frac{1}{2}
\end{cases}, & \gamma_{-,min} & =\begin{cases}
1 & \text{if }\gamma_{+}\geq\gamma_{B}\left(\varepsilon\right)\text{ and }\varepsilon\geq\frac{1}{2}\\
\gamma^{+}\left(\varepsilon,\gamma_{+}\right) & \text{if }\gamma_{+}<\gamma_{B}\left(\varepsilon\right)\text{ and }\varepsilon\geq1\\
\gamma^{-}\left(\varepsilon,\gamma_{+}\right) & \text{for all other cases}
\end{cases},
\end{align}

The angle averaged emissivity is given by:
\begin{equation}
\overline{v\frac{d\sigma}{d\varepsilon}}\left(\varepsilon,\gamma_{+},\gamma_{-}\right)=\pi cr_{e}^{2}\frac{1}{\beta_{+}\beta_{-}\gamma_{+}^{2}\gamma_{-}^{2}}\left[\sqrt{\left(\gamma_{+}+\gamma_{-}\right)^{2}-4\gamma_{cm}^{2}}+H_{+}+H_{-}\right]_{\gamma_{cm,min}}^{\gamma_{cm.max}},
\end{equation}
where the limits are given by kinetic reasoning:
\begin{align}
\gamma_{cm}^{\pm2} & =\frac{1}{2}\left(1+\gamma_{+}\gamma_{-}\pm\beta_{+}\beta_{-}\gamma_{+}\gamma_{-}\right),\\
\gamma_{cm,min} & =\gamma_{cm}^{-}, & \gamma_{cm,max} & =\min\left(\gamma_{cm}^{+},\gamma_{cm}^{*}\right),
\end{align}

The $H_{\pm}$ is given again in a form that depends on whether $c_{\pm}$
is equal or different from $0$.
\begin{equation}
H_{\pm}=\begin{cases}
\left(2+\frac{1-\gamma_{cm}^{*2}}{c_{\pm}}\right)I_{\pm}+\frac{1}{u_{\pm}}\left[\frac{1}{\gamma_{cm}}-\frac{\gamma_{cm}}{c_{\pm}}+\frac{\gamma_{cm}}{2\gamma_{cm}^{*2}}\left(2c_{\pm}-d_{\pm}\right)\right]+u_{\pm}\frac{\gamma_{cm}}{c_{\pm}} & \text{if }c_{\pm}\neq0\\
\left(\frac{2}{3}\gamma_{cm}^{3}+2\gamma_{cm}+\frac{1}{\gamma_{cm}}\right)\frac{1}{\gamma_{cm}^{*}}+\frac{1}{2}\left(\frac{2}{3}\gamma_{cm}^{3}-d_{\pm}\gamma_{cm}\right)\frac{1}{\gamma_{cm}^{*3}} & \text{if }c_{\pm}=0
\end{cases},
\end{equation}
where
\begin{equation}
I_{\pm}=\begin{cases}
\frac{1}{\sqrt{c_{\pm}}}\ln\left(\gamma_{cm}\sqrt{c_{\pm}}+u_{\pm}\right) & \text{if }c_{\pm}>0\\
\frac{1}{\sqrt{-c_{\pm}}}\arcsin\left(\frac{\gamma_{cm}}{\gamma_{cm}^{*}}\sqrt{-c_{\pm}}\right) & \text{if }c_{\pm}<0
\end{cases}.
\end{equation}

\subsection{Photo-Meson Production}

Hadrons (protons and neutrons) interact with photons producing hadronic
resonances which ultimately decay in the form of hadrons and pions.
To model this we follow the approximations of Hümmer et al \citep{hummer_simplified_2010}
in their SimB model. We diverge though in the calculations of the
loss and self-injection term for hadrons as, instead of applying a
cooling term, we model every process as a loss and where applicable,
we reinject the corresponding hadron with reduced energies.

For this process, and in order to avoid adding constants and units
we choose to follow the original convention on units and notation.
We begin with the equation for the injection of particles from an
interaction between a photon and a proton:
\begin{equation}
\mathcal{Q}_{b}^{IT}=N_{p}\left(\frac{E_{b}}{\chi^{IT}}\right)\frac{m_{p}}{E_{b}}\int_{\epsilon_{th}/2}^{\infty}dyn_{\gamma}\left(\frac{m_{p}y\chi^{IT}}{E_{b}}\right)M_{b}^{IT}f^{IT}\left(y\right),
\end{equation}
where $IT$ makes reference to the Interaction Type and $b$ to the
daughter particle. The constants $\chi^{IT}$, $M_{b}^{IT}$ and $\epsilon_{th}$
and the function $f^{IT}\left(y\right)$ depend on the interaction
type.

\bigskip{}

For the Resonant Pion production, we follow their Simplified Model
B \citep[section 4.2.2 \& Table 4]{hummer_simplified_2010}, whose
parameters we reproduce in \tabref{Sim-B-parameters-for-resonant-pion-production}.

\begin{table}
\begin{centering}
\begin{tabular}{cccccccccccc}
\toprule 
\multirow{3}{*}{IT} & \multirow{3}{*}{$\epsilon_{r}$ Range (GeV)} & \multirow{3}{*}{$\sigma$($\mu$barn)} & \multicolumn{4}{c}{Initial Proton} & \multicolumn{4}{c}{Initial Neutron} & \multirow{3}{*}{K}\tabularnewline
\cmidrule{4-11} \cmidrule{5-11} \cmidrule{6-11} \cmidrule{7-11} \cmidrule{8-11} \cmidrule{9-11} \cmidrule{10-11} \cmidrule{11-11} 
 &  &  & $M_{\pi^{0}}$ & $M_{\pi^{+}}$ & $M_{\pi^{-}}$ & $M_{n}$ & $M_{\pi^{0}}$ & $M_{\pi^{+}}$ & $M_{\pi^{-}}$ & $M_{n}$ & \tabularnewline
\cmidrule{4-11} \cmidrule{5-11} \cmidrule{6-11} \cmidrule{7-11} \cmidrule{8-11} \cmidrule{9-11} \cmidrule{10-11} \cmidrule{11-11} 
 &  &  & $\chi_{\pi^{0}}$ & $\chi_{\pi^{+}}$ & $\chi_{\pi^{-}}$ & $M_{p}$ & $\chi_{\pi^{0}}$ & $\chi_{\pi^{+}}$ & $\chi_{\pi^{-}}$ & $M_{p}$ & \tabularnewline
\midrule 
\multirow{2}{*}{LR} & \multirow{2}{*}{$0.2-0.5$} & \multirow{2}{*}{$200$} & $2/3$ & $1/3$ &  & $1/3$ & $2/3$ &  & $1/3$ & $2/3$ & \multirow{2}{*}{$0.22$}\tabularnewline
\cmidrule{4-11} \cmidrule{5-11} \cmidrule{6-11} \cmidrule{7-11} \cmidrule{8-11} \cmidrule{9-11} \cmidrule{10-11} \cmidrule{11-11} 
 &  &  & $0.22$ & $0.22$ &  & $2/3$ & $0.22$ &  & $0.22$ & $1/3$ & \tabularnewline
\midrule 
\multirow{2}{*}{HR} & \multirow{2}{*}{$0.5-1.2$} & \multirow{2}{*}{$90$} & $0.47$ & $0.77$ & $0.34$ & $0.43$ & $0.47$ & $0.34$ & $0.77$ & $0.57$ & \multirow{2}{*}{$0.39$}\tabularnewline
\cmidrule{4-11} \cmidrule{5-11} \cmidrule{6-11} \cmidrule{7-11} \cmidrule{8-11} \cmidrule{9-11} \cmidrule{10-11} \cmidrule{11-11} 
 &  &  & $0.26$ & $0.25$ & $0.22$ & $0.57$ & $0.26$ & $0.22$ & $0.25$ & $0.43$ & \tabularnewline
\bottomrule
\end{tabular}
\par\end{centering}
\caption{\label{tab:Sim-B-parameters-for-resonant-pion-production}Sim-B parameters
for the resonant pion production.}

\end{table}

The $f$ functions are given depending on the Interaction Type as:
\begin{equation}
f^{LR}=\begin{cases}
0 & \text{if }2y<0.2\text{GeV}\\
200\mu\text{barn}\left(1-\frac{\left(0.2\text{GeV}\right)^{2}}{\left(2y\right)^{2}}\right) & \text{if }0.2\text{GeV}\leq2y<0.5\text{GeV}\\
200\mu\text{barn}\left(\frac{\left(0.5\text{GeV}\right)^{2}-\left(0.2\text{GeV}\right)^{2}}{\left(2y\right)^{2}}\right) & \text{if }0.5\text{GeV}\leq2y
\end{cases},
\end{equation}
\begin{equation}
f^{HR}=\begin{cases}
0 & \text{if }2y<0.5\text{GeV}\\
90\mu\text{barn}\left(1-\frac{\left(0.5\text{GeV}\right)^{2}}{\left(2y\right)^{2}}\right) & \text{if }0.5\text{GeV}\leq2y<1.2\text{GeV}\\
90\mu\text{barn}\left(\frac{\left(1.2\text{GeV}\right)^{2}-\left(0.5\text{GeV}\right)^{2}}{\left(2y\right)^{2}}\right) & \text{if }1.2\text{GeV}\leq2y
\end{cases}
\end{equation}
\bigskip{}

For the direct pion production, we follow \citep[section 4.3 \& Table 5]{hummer_simplified_2010},
whose parameters we reproduce in \tabref{Parameters-for-the-direct-pion-production}.

\begin{table}
\begin{centering}
\begin{tabular}{ccccccccccc}
\toprule 
\multirow{2}{*}{IT} & \multirow{2}{*}{$\epsilon_{min}^{IT}$ (GeV)} & \multirow{2}{*}{$\epsilon_{max}^{IT}$ (GeV)} & \multirow{2}{*}{$\chi$} & \multirow{2}{*}{K} & \multicolumn{3}{c}{Initial Proton} & \multicolumn{3}{c}{Initial Neutron}\tabularnewline
\cmidrule{6-11} \cmidrule{7-11} \cmidrule{8-11} \cmidrule{9-11} \cmidrule{10-11} \cmidrule{11-11} 
 &  &  &  &  & $M_{\pi^{0}}$ & $M_{\pi^{+}}$ & $M_{\pi^{-}}$ & $M_{\pi^{0}}$ & $M_{\pi^{+}}$ & $M_{\pi^{-}}$\tabularnewline
\midrule 
$T_{1L}$ & $0.17$ & $0.56$ & $0.13$ & $0.13$ &  & $1$ &  &  &  & $1$\tabularnewline
$T_{1M}$ & $0.56$ & $10$ & $0.05$ & $0.05$ &  & $1$ &  &  &  & $1$\tabularnewline
$T_{1H}$ & $10$ & $\infty$ & $0.001$ & $0.001$ &  & $1$ &  &  &  & $1$\tabularnewline
\midrule 
$T_{2aL}$ & $0.4$ & $1.58$ & $0.08$ & $0.28$ &  & $1/4$ & $3/4$ &  & $3/4$ & $1/4$\tabularnewline
$T_{2aM}$ & $1.58$ & $10$ & $0.02$ & $0.22$ &  & $1/4$ & $3/4$ &  & $3/4$ & $1/4$\tabularnewline
$T_{2aH}$ & $10$ & $\infty$ & $0.001$ & $0.201$ &  & $1/4$ & $3/4$ &  & $3/4$ & $1/4$\tabularnewline
\midrule 
$T_{2b}$ & $0.4$ & $\infty$ & $0.2$ &  & $1/6$ & $3/4$ & $1/12$ & $1/6$ & $1/12$ & $3/4$\tabularnewline
\bottomrule
\end{tabular}
\par\end{centering}
\caption{\label{tab:Parameters-for-the-direct-pion-production}Parameters for
the direct pion production}

\end{table}

The $f$ functions are given by;
\begin{equation}
f^{IT}=\begin{cases}
0 & \text{if }2y<\epsilon_{min}^{IT}\\
\frac{1}{2y^{2}}\left(I^{IT}\left(2y\right)-I^{IT}\left(\epsilon_{min}^{IT}\right)\right) & \text{if }\epsilon_{min}^{IT}\leq2y<\epsilon_{max}^{IT}\\
\frac{1}{2y^{2}}\left(I^{IT}\left(\epsilon_{max}^{IT}\right)-I^{IT}\left(\epsilon_{min}^{IT}\right)\right) & \text{if }\epsilon_{max}^{IT}\leq2y
\end{cases}.
\end{equation}
The functions $I^{IT}$ are parametrized in terms of $x\equiv\log_{10}\left(y\text{(GeV)}\right)$
and are different depending on the interaction type:
\begin{equation}
I^{T_{1}}=\begin{cases}
0 & \text{if }2y<0.17\text{GeV}\\
35.9533+84.0859x+110.765x^{2}+102.728x^{3}+40.4699x^{4} & \text{if }0.17\text{GeV}<2y<0.96\text{GeV}\\
30.2004+40.5478x+2.0374x^{2}-0.387884x^{3}+0.025044x^{4} & \text{if }0.96\text{GeV}<2y
\end{cases},
\end{equation}

\begin{equation}
I^{T_{2}}=\begin{cases}
0 & \text{if }2y<0.4\text{GeV}\\
-3.4083+\frac{16.2864}{2y}+40.7160\ln\left(2y\right) & \text{if }0.4\text{GeV}<2y
\end{cases}.
\end{equation}
\bigskip{}

For the multi pion production, we follow \citep[section 4.4.2 \& Table 6]{hummer_simplified_2010},
whose parameters we reproduce in \tabref{Parameters-for-the-multi-pion-production}.

\begin{table}
\begin{centering}
\begin{tabular}{cccccccccccc}
\toprule 
\multirow{2}{*}{IT} & \multirow{2}{*}{$\epsilon_{min}^{IT}$ (GeV)} & \multirow{2}{*}{$\epsilon_{max}^{IT}$ (GeV)} & \multirow{2}{*}{$\sigma$} & \multirow{2}{*}{$\chi$} & \multirow{2}{*}{K} & \multicolumn{3}{c}{Initial Proton} & \multicolumn{3}{c}{Initial Neutron}\tabularnewline
\cmidrule{7-12} \cmidrule{8-12} \cmidrule{9-12} \cmidrule{10-12} \cmidrule{11-12} \cmidrule{12-12} 
 &  &  &  &  &  & $M_{\pi^{0}}$ & $M_{\pi^{+}}$ & $M_{\pi^{-}}$ & $M_{\pi^{0}}$ & $M_{\pi^{+}}$ & $M_{\pi^{-}}$\tabularnewline
\midrule 
$M_{1L}$ & 0.5 & 0.9 & 60 & 0.1 & 0.27 & 0.32 & 0.34 & 0.04 & 0.32 & 0.04 & 0.34\tabularnewline
$M_{1H}$ & 0.5 & 0.9 & 60 & 0.4 &  & 0.17 & 0.29 & 0.05 & 0.17 & 0.05 & 0.29\tabularnewline
$M_{2L}$ & 0.9 & 1.5 & 85 & 0.15 & 0.34 & 0.42 & 0.31 & 0.07 & 0.42 & 0.07 & 0.31\tabularnewline
$M_{2H}$ & 0.9 & 1.5 & 85 & 0.35 &  & 0.19 & 0.35 & 0.08 & 0.19 & 0.08 & 0.35\tabularnewline
$M_{3L}$ & 1.5 & 5.0 & 120 & 0.15 & 0.39 & 0.59 & 0.57 & 0.30 & 0.59 & 0.30 & 0.57\tabularnewline
$M_{3H}$ & 1.5 & 5.0 & 120 & 0.35 &  & 0.16 & 0.21 & 0.13 & 0.16 & 0.13 & 0.21\tabularnewline
$M_{4L}$ & 5.0 & 50 & 120 & 0.07 & 0.49 & 1.38 & 1.37 & 1.11 & 1.38 & 1.11 & 1.37\tabularnewline
$M_{4H}$ & 5.0 & 50 & 120 & 0.35 &  & 0.16 & 0.25 & 0.23 & 0.16 & 0.23 & 0.25\tabularnewline
$M_{5L}$ & 50 & 500 & 120 & 0.02 & 0.45 & 3.01 & 2.86 & 2.64 & 3.01 & 2.64 & 2.86\tabularnewline
$M_{5H}$ & 50 & 500 & 120 & 0.5 &  & 0.20 & 0.21 & 0.14 & 0.20 & 0.14 & 0.21\tabularnewline
$M_{6L}$ & 500 & 5000 & 120 & 0.007 & 0.44 & 5.13 & 4.68 & 4.57 & 5.13 & 4.57 & 4.68\tabularnewline
$M_{6H}$ & 500 & 5000 & 120 & 0.5 &  & 0.27 & 0.29 & 0.12 & 0.27 & 0.12 & 0.29\tabularnewline
$M_{7L}$ & 5000 & $\infty$ & 120 & 0.002 & 0.44 & 7.59 & 6.80 & 6.65 & 7.59 & 6.65 & 6.80\tabularnewline
$M_{7H}$ & 5000 & $\infty$ & 120 & 0.6 &  & 0.26 & 0.27 & 0.13 & 0.26 & 0.13 & 0.27\tabularnewline
\bottomrule
\end{tabular}
\par\end{centering}
\caption{\label{tab:Parameters-for-the-multi-pion-production}Parameters for
the multi pion production.}
\end{table}

The $f$ functions are given by;
\begin{equation}
f^{IT}=\begin{cases}
0 & \text{if }2y<\epsilon_{min}^{M_{i}}\\
\frac{\sigma^{M_{i}}}{\left(2y\right)^{2}}\left(\left(2y\right)^{2}-\left(\epsilon_{min}^{M_{i}}\right)^{2}\right) & \text{if }\epsilon_{min}^{M_{i}}\leq2y<\epsilon_{max}^{M_{i}}\\
\frac{\sigma^{M_{i}}}{\left(2y\right)^{2}}\left(\left(\epsilon_{max}^{M_{i}}\right)^{2}-\left(\epsilon_{min}^{M_{i}}\right)^{2}\right) & \text{if }\epsilon_{max}^{M_{i}}\leq2y
\end{cases}.
\end{equation}
\bigskip{}

To account for energy losses in protons and neutrons, we also follow
\citet[Appendix B]{hummer_simplified_2010}, although we apply the
framework of reinjection to all the particles and we ignore cooling.
This way, all the losses can be treated with the same equation:
\begin{equation}
\mathcal{Q}_{p'}^{IT}=N_{p}\left(\frac{E_{p'}}{1-K^{IT}}\right)\frac{m_{p}}{E_{p'}}\int_{\epsilon_{th}/2}^{\infty}dyn_{\gamma}\left(\frac{m_{p}y\left(1-K^{IT}\right)}{E_{p'}}\right)M_{p'}^{IT}f^{IT}\left(y\right),
\end{equation}
where $p'$ is the resulting hadron and $K^{IT}$ is the inelasticity
of the interaction type and the corresponding channel.

\subsection{Bethe-Heitler Process}

Besides producing mesons, a photon interacting with a proton can produce
pairs electron-positron through the Bethe-Heitler process. We have
chosen to model this process using the results of \citet[Section IV]{kelner_energy_2008}.

First, the injection of produced particles is defined as:
\begin{equation}
\frac{dn_{e^{\pm}}\left(\gamma_{e}\right)}{d\gamma_{e}}=\frac{c}{2\gamma_{p}^{3}}\int_{\frac{\left(\gamma_{p}+\gamma_{e}\right)^{2}}{4\gamma_{p}^{2}\gamma_{e}}}^{\infty}d\varepsilon\frac{n_{\gamma}\left(\varepsilon\right)}{\varepsilon^{2}}\int_{\frac{\left(\gamma_{p}+\gamma_{e}\right)^{2}}{2\gamma_{p}\gamma_{e}}}^{2\gamma_{p}\varepsilon}d\omega\omega\int_{\frac{\gamma_{p}^{2}+\gamma_{e}^{2}}{2\gamma_{p}\gamma_{e}}}^{\omega-1}\frac{d\gamma_{-}}{p_{-}}W\left(\omega,\gamma_{-},\xi\right),\label{eq:BH_dne/dgp}
\end{equation}
where $\gamma_{e}$ is the energy of the final electron (or correspondingly
positron), $\gamma_{p}$ is the energy of the initial proton and $\varepsilon$
is the energy of the incoming photon. $\omega$ and $\gamma_{-}$
act as integration variables which act as energies in the rest frame
of the proton. Note that we have added a factor $c$ with respect
to the original equation to correct the units.

In the limit of highly energetic protons we have that:
\begin{equation}
\cos\theta=\xi=\frac{\gamma_{p}\gamma_{-}-\gamma_{e}}{\gamma_{p}\beta_{-}\gamma_{-}}.
\end{equation}

$W$ is the differential reaction rate:
\begin{equation}
W\left(\omega,\gamma_{-},\xi\right)=\frac{d\sigma^{2}}{d\gamma_{-}d\cos\theta_{-}},
\end{equation}
which is available in \citet[equation 10]{blumenthal_energy_1970}
under de assumption of having a proton with infinite mass:
\begin{align}
\frac{d\sigma^{2}}{d\gamma_{-}d\cos\theta_{-}} & =\left(\frac{\alpha Z^{2}r_{0}^{2}p_{-}p_{+}}{2\omega^{3}}\right)\left\{ -4\sin^{2}\theta_{-}\frac{2\gamma_{-}^{2}+1}{p_{-}^{2}\Delta_{-}^{4}}+\frac{5\gamma_{-}^{2}-2\gamma_{+}\gamma_{-}+3}{p_{-}^{2}\Delta_{-}^{2}}+\frac{p_{-}^{2}-\omega^{2}}{T^{2}\Delta_{-}^{2}}+\frac{2\gamma_{+}}{p_{-}^{2}\Delta_{-}}+\right.\nonumber \\
 & +\frac{Y}{p_{-}p_{+}}\left[2\gamma_{-}\sin^{2}\theta_{-}\frac{3\omega+p_{-}^{2}\gamma_{+}}{\Delta_{-}^{4}}+\frac{2\gamma_{-}^{2}\left(\gamma_{-}^{2}+\gamma_{+}^{2}\right)-7\gamma_{-}^{2}-3\gamma_{+}\gamma_{-}-\gamma_{+}^{2}+1}{\Delta_{-}^{2}}+\frac{\omega\left(\gamma_{-}^{2}-\gamma_{-}\gamma_{+}-1\right)}{\Delta_{-}}\right]\\
 & \left.-\frac{\delta_{+}^{T}}{p_{+}T}\left[\frac{2}{\Delta_{-}^{2}}-\frac{3\omega}{\Delta_{-}}-\frac{\omega\left(p_{-}^{2}-\omega^{2}\right)}{T^{2}\Delta_{-}}\right]-\frac{2y_{+}}{\Delta_{-}}\right\} ,\nonumber 
\end{align}
where
\begin{align}
T & =\sqrt{p_{+}^{2}+2\omega\Delta_{-}}, & \delta_{+}^{T} & =\ln\left(1+p_{+}\frac{p_{+}+T}{\omega\Delta_{-}}\right),\\
Y & =\frac{2}{p_{-}^{2}}\ln\left(\frac{\gamma_{-}\gamma_{+}+p_{-}p_{+}+1}{\omega}\right), & \Delta_{-} & =\frac{\gamma_{e}}{\gamma_{p}},\\
y_{+} & =\frac{2}{p_{+}}\atanh\left(\beta_{+}\right),
\end{align}
and $\alpha$ is the fine structure constant, $Z$ is the electric
charge of the scattering nucleus (in our case for protons $Z=1$)
and $r_{0}$ is the classical radius of the electron. We have chosen
to keep the notation compact $p_{\pm}=\beta_{\pm}\gamma_{\pm}$ to
denote the modulus of the momentum of a particle. Under the assumption
of the proton having infinite mass, or being at rest, we have that
$\omega=\gamma_{-}+\gamma_{+}$.

We will focus on the second and third integrals, which only depend
on external variables:
\begin{equation}
F\left(\gamma_{p},\gamma_{e},\varepsilon\right)=\int_{\frac{1}{2}\left(\frac{1}{\Delta_{-}}+\Delta_{-}\right)+1}^{2\gamma_{p}\varepsilon}d\omega\omega\int_{\frac{1}{2}\left(\frac{1}{\Delta_{-}}+\Delta_{-}\right)}^{\omega-1}\frac{d\gamma_{-}}{p_{-}}W\left(\omega,\gamma_{-},\Delta_{-}\right).
\end{equation}

Pulling out some constants we obtain:
\begin{align}
F\left(\gamma_{p},\gamma_{e},\varepsilon\right) & =\frac{\alpha Z^{2}r_{0}^{2}}{2\Delta_{-}}\int_{\frac{1}{2}\left(\frac{1}{\Delta_{-}}+\Delta_{-}\right)+1}^{2\gamma_{p}\varepsilon}\frac{d\omega}{\omega^{2}}G'\left(\omega,\Delta_{-}\right),\\
G'\left(\omega,\Delta_{-}\right) & =\int_{\frac{1}{2}\left(\frac{1}{\Delta_{-}}+\Delta_{-}\right)}^{\omega-1}dE_{-}W'\left(\omega,E_{-},\Delta_{-}\right),\\
W_{1}'\left(\omega,\gamma_{-},\Delta_{-}\right) & =p_{+}\left(-4\sin^{2}\theta_{-}\frac{2\gamma_{-}^{2}+1}{p_{-}^{2}\Delta_{-}^{3}}+\frac{5\gamma_{-}^{2}-2\gamma_{+}\gamma_{-}+3}{p_{-}^{2}\Delta_{-}}+\frac{p_{-}^{2}-\omega^{2}}{T^{2}\Delta_{-}}+\frac{2\gamma_{+}}{p_{-}^{2}}\right),\\
W_{2}'\left(\omega,\gamma_{-},\Delta_{-}\right) & =\frac{Y}{p_{-}}\left[2\gamma_{-}\sin^{2}\theta_{-}\frac{3\omega+p_{-}^{2}\gamma_{+}}{\Delta_{-}^{3}}+\frac{2\gamma_{-}^{2}\left(\gamma_{-}^{2}+\gamma_{+}^{2}\right)-7\gamma_{-}^{2}-3\gamma_{+}\gamma_{-}-\gamma_{+}^{2}+1}{\Delta_{-}}+\omega\left(\gamma_{-}^{2}-\gamma_{-}\gamma_{+}-1\right)\right],\\
W_{3}'\left(\omega,\gamma_{-},\Delta_{-}\right) & =-\frac{\delta_{+}^{T}}{T}\left[\frac{2}{\Delta_{-}}-3\omega-\frac{\omega\left(p_{-}^{2}-\omega^{2}\right)}{T^{2}}\right],\\
W_{4}'\left(\omega,\gamma_{-},\Delta_{-}\right) & =-4\atanh\left(\beta_{+}\right).
\end{align}

As far as we know, no simple form for these integrals exist so they
are performed numerically using an adaptive integration procedure
from the Gnu Scientific Library.

Finally, to calculate the injection rate for electrons or positrons,
\eqref{BH_dne/dgp} has to be integrated over the proton population:
\begin{equation}
\mathcal{Q}_{e}\left(\gamma_{e}\right)=\int_{1}^{\infty}\frac{dn_{e^{\pm}}\left(\gamma_{e}\right)}{d\gamma_{e}}n_{p}\left(\gamma_{p}\right)d\gamma_{p}.
\end{equation}

\subsection{Pion-Muon Decays}

To account for the cascade of pion and muon decays we follow the same
model of \citet{lipari_flavor_2007} and \citet{hummer_simplified_2010}.

The injection of produced particles can be factorized as:
\begin{equation}
Q_{b}\left(E_{b}\right)=\int dE_{a}\frac{N_{a}\left(E_{a}\right)}{\tau_{a}}\frac{F_{a\rightarrow b}\left(E_{b}/E_{a}\right)}{E_{a}},
\end{equation}
where $E_{a}$ and $E_{b}$ are the energies of the parent and daughter
particles (not normalized), $\tau_{a}$ is the decay timescale of
the parent particle and $F_{a\rightarrow b}$ is a function that gives
the ratio of decaying.

Charged pions decay to muons, which subsequently decay through the
following channels:

\begin{align*}
\pi^{+}\rightarrow & \mu^{+}+\nu_{\mu}\\
 & \mu^{+}\rightarrow e^{+}+\overline{\nu_{\mu}}+\nu_{e}\\
\pi^{-}\rightarrow & \mu^{-}+\overline{\nu_{\mu}}\\
 & \mu^{-}\rightarrow e^{-}+\overline{\nu_{\mu}}+\nu_{e}
\end{align*}

The rate of charged pions decaying into neutrinos is given by \citet{lipari_flavor_2007}:
\begin{equation}
F_{\pi^{+}\rightarrow\nu_{\mu}}\left(x\right)=F_{\pi^{-}\rightarrow\overline{\nu_{\mu}}}\left(x\right)=\frac{1}{1-r_{\pi}}\Theta\left(1-r_{\pi}-x\right).
\end{equation}

For the decay into muons we need to track the helicity of the final
muons, as it will impact the subsequent decays into neutrinos, \citet{hummer_simplified_2010}:
\begin{align}
F_{\pi^{+}\rightarrow\mu_{R}^{+}}\left(x\right) & =F_{\pi^{-}\rightarrow\mu_{L}^{-}}\left(x\right)=\frac{\left(1-x\right)r_{\pi}}{x\left(1-r_{\pi}\right)^{2}}\Theta\left(x-r_{\pi}\right),\\
F_{\pi^{+}\rightarrow\mu_{L}^{+}}\left(x\right) & =F_{\pi^{-}\rightarrow\mu_{R}^{-}}\left(x\right)=\frac{x-r_{\pi}}{x\left(1-r_{\pi}\right)^{2}}\Theta\left(x-r_{\pi}\right).
\end{align}

Then, decay of muons to neutrinos is given by \citet{hummer_simplified_2010}:
\begin{align}
F_{\mu^{+}\rightarrow\overline{\nu_{\mu}}}\left(x,h\right) & =F_{\mu^{-}\rightarrow\nu_{\mu}}\left(x,-h\right)=\left(\frac{5}{3}+3x^{2}+\frac{3}{4}x^{3}\right)+h\left(-\frac{1}{3}+3x^{2}-\frac{8}{3}x^{3}\right),\\
F_{\mu^{+}\rightarrow\nu_{e}}\left(x,h\right) & =F_{\mu^{-}\rightarrow\overline{\nu_{e}}}\left(x,-h\right)=\left(2-6x^{2}+4x^{3}\right)+h\left(2-12x+18x^{2}-8x^{3}\right),
\end{align}
with $h=1$ for right handed muons and $h=-1$ for left handed muons.

The functions that govern the subsequent injection of electrons and
positrons do depend on the helicity of the parent muons and can be
simplified to:
\begin{align}
F_{\mu_{L}^{+}\rightarrow e^{+}} & \left(x\right)=F_{\mu_{R}^{-}\rightarrow e^{-}}\left(x\right)=2-3x^{2}+6x^{3},\\
F_{\mu_{R}^{+}\rightarrow e^{+}} & \left(x\right)=F_{\mu_{L}^{-}\rightarrow e^{-}}\left(x\right)=\frac{4}{3}-\frac{4}{2}x^{3}.
\end{align}

\subsection{Neutral Pion Decay}

The injection of photons from neutral pions follows the simple equation:
\begin{equation}
\mathcal{Q}_{\gamma}\left(\varepsilon\right)=\int_{\min\left(1,\frac{m_{e}}{m_{\pi}}\varepsilon\right)}\frac{n_{\pi^{0}\left(\gamma_{\pi^{0}}\right)}}{\gamma_{\pi^{0}}\tau_{\pi^{0}}}d\gamma_{\pi^{0}}.
\end{equation}

\twocolumn

\section{Configuration File}{\label{sec:Configuration-File}}

\texttt{Katu} is very easy to configure, this is done is done with
the \texttt{Toml}\footnote{\texttt{\href{https://toml.io/en/}{https://toml.io/en/}}}
language, which allows us to separate variables per tables and subtables.
An example configuration can be found in the \texttt{default.toml}
file. As a reference, we detail all the possible options here.

The \texttt{{[}general{]}} table refers to general options for the
simulation and the plasma. The possible keys are:
\begin{itemize}
\item \texttt{density}: The density of the background plasma
\item \texttt{magnetic\_field}: The magnitude of the background magnetic
field
\item \texttt{eta}: The ratio of protons to electrons in the background
plasma
\item \texttt{cfe\_ratio}: The ratio of charged to free escape
\item \texttt{t\_acc}: The acceleration timescale
\item \texttt{dt}: The initial value of the time step
\item \texttt{dt\_max}: The maximum value of the time step
\item \texttt{t\_max}: The maximum simulation time
\item \texttt{tol}: The requested tolerance for the steady state
\end{itemize}
The \texttt{{[}volume{]}} table refers to options referred to the
shape of the simulation volume.
\begin{itemize}
\item \texttt{shape}: The shape of the volume, currently it supports \texttt{\textquotedbl sphere\textquotedbl}
and \texttt{\textquotedbl disk\textquotedbl}
\item \texttt{R}: The radius of the volume
\item \texttt{h}: The height of the disk. (Only used for the \texttt{\textquotedbl disk\textquotedbl}
shape)
\end{itemize}
The \texttt{external\_injection} table refers to the injection of
particles from outside the simulated volume. The possible keys are:
\begin{itemize}
\item \texttt{luminosity}: The integrated luminosity of the injected particles
\item \texttt{eta}: The ratio of protons to electrons to inject.
\end{itemize}
Note that the subtable \texttt{external\_injection.photons} also has
\texttt{luminosity} as a key separated from the particle injection.

The distribution tables and subtables (\texttt{{[}protons{]}}, \texttt{{[}electrons{]}}
and \texttt{{[}photons{]}}) refer to the distribution of particles,
their limits and number of bins. The available keys are:
\begin{itemize}
\item \texttt{gamma\_min}: The minimum value for gamma for protons and electrons
\item \texttt{gamma\_max}: The maximum value for gamma for protons and electrons
\item \texttt{epsilon\_min}: The minimum value for epsilon for photons
\item \texttt{epsilon\_max}: The maximum value for epsilon for photons
\item \texttt{size}: The number of bins for the particle kind
\end{itemize}
The distribution of the corresponding particle is controlled by the
key \texttt{distribution\_type} and supports the following options,
along with the corresponding keys:
\begin{itemize}
\item \texttt{\textquotedbl maxwell\_juttner\textquotedbl}: This option
gives a thermal distribution of particles
\begin{itemize}
\item \texttt{temperature}: The temperature of the distribution, in terms
of the rest mass energy of the particle
\end{itemize}
\item \texttt{\textquotedbl black\_body\textquotedbl}: This option gives
a thermal distribution of photons
\begin{itemize}
\item \texttt{temperature}: The temperature of the distribution, in terms
of the rest mass energy of an electron
\end{itemize}
\item \texttt{\textquotedbl power\_law\textquotedbl}: This option gives
a simple power law
\begin{itemize}
\item \texttt{slope}: The value of the exponent of the power law
\end{itemize}
\item \texttt{\textquotedbl broken\_power\_law\textquotedbl}: This option
gives a broken power law
\begin{itemize}
\item \texttt{break\_point}: The value of the energy where the the power
law changes from the first slope to the second
\item \texttt{first\_slope}: The value of the first exponent of the power
law
\item \texttt{second\_slope}: The value of the second exponent of the power
law
\end{itemize}
\item \texttt{\textquotedbl connected\_power\_law\textquotedbl}: This
option gives a smooth power law with two slopes
\begin{itemize}
\item \texttt{connection\_point}: The value of the energy where the the
power law changes from the first slope to the second
\item \texttt{first\_slope}: The value of the first exponent of the power
law
\item \texttt{second\_slope}: The value of the second exponent of the power
law
\end{itemize}
\item \texttt{\textquotedbl power\_law\_with\_exponential\_cutoff\textquotedbl}:
This option gives a power law with an exponential cutoff
\begin{itemize}
\item \texttt{slope}: The value of the exponent of the power law
\item \texttt{break\_point}: The normalization value for the energy in the
exponential cutoff
\end{itemize}
\item \texttt{\textquotedbl hybrid\textquotedbl}: This option gives a
thermal distribution of particles at low energies coupled with a power
law at higher energies
\begin{itemize}
\item \texttt{temperature}: The temperature of the distribution, in terms
of the rest mass energy
\item \texttt{slope}: The value of the exponent of the power law
\end{itemize}
\end{itemize}

\section{Corner Plots}

\begin{figure*}
\begin{centering}
\includegraphics[width=0.9\paperwidth]{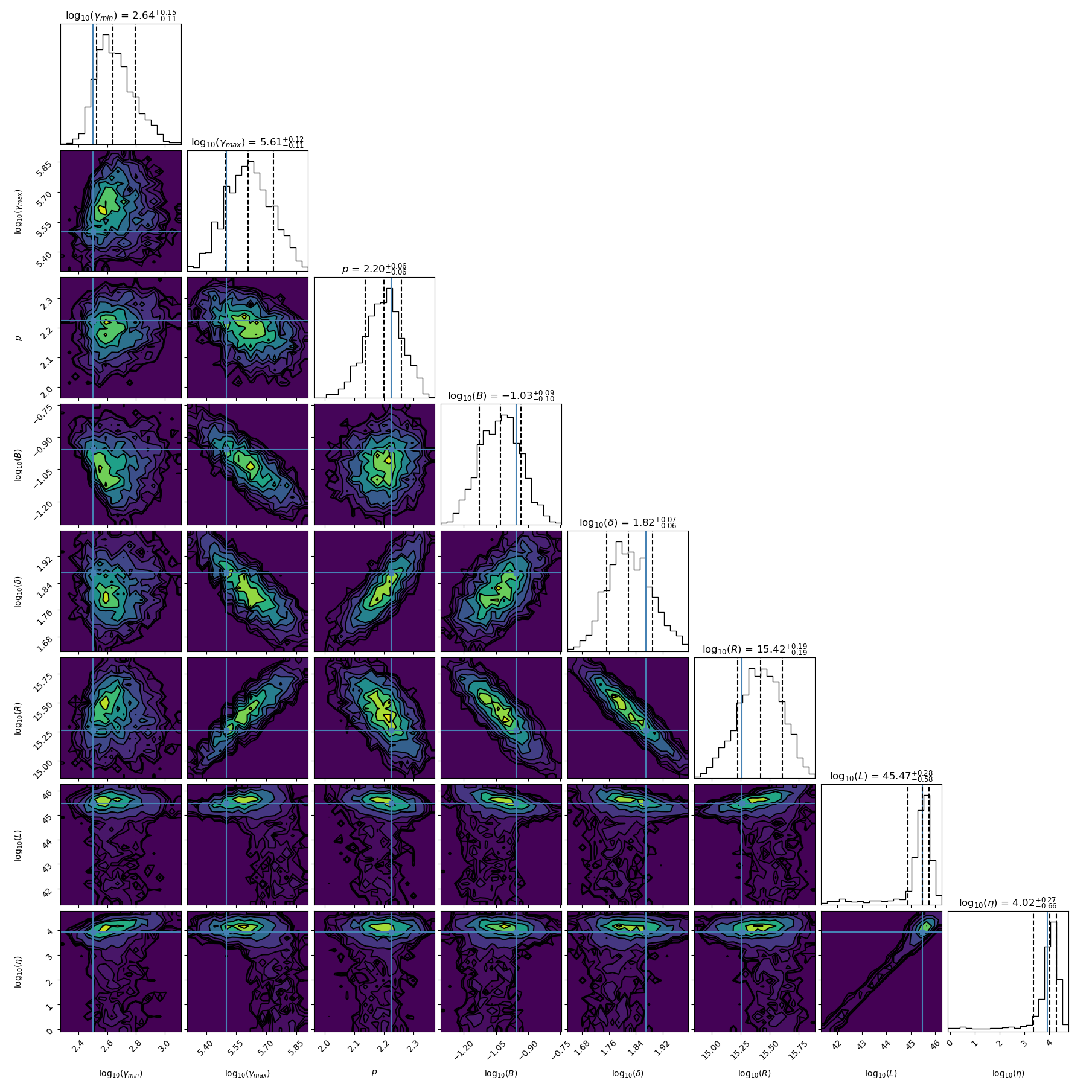}
\par\end{centering}
\caption{\label{fig:Corner-Plot-multinest}Corner Plot for \texttt{multinest}.
The blue line represents the best loglikelihood that has been found.}
\end{figure*}

\begin{figure*}
\begin{centering}
\includegraphics[width=0.9\paperwidth]{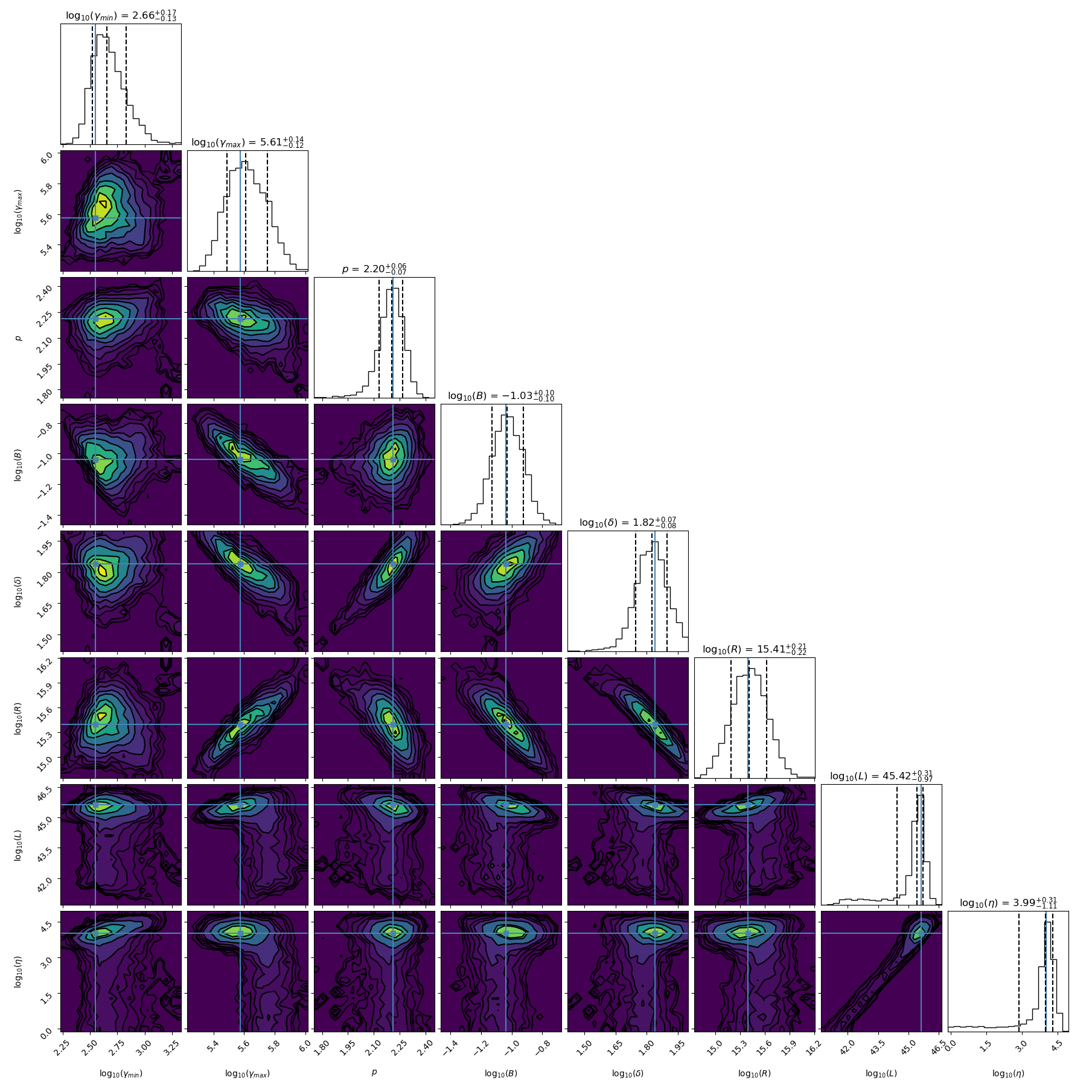}
\par\end{centering}
\caption{\label{fig:Corner-Plot-emcee}Corner Plot for \texttt{emcee}. The
blue line represents the best loglikelihood that has been found.}
\end{figure*}

\bsp 

\label{lastpage} 
\end{document}